%% file: main.tex
\documentclass[lettersize,journal]{IEEEtran}
\usepackage{amsmath,amsfonts}
\usepackage{algorithmic}
\usepackage{array}
\usepackage[caption=false,font=normalsize,labelfont=sf,textfont=sf]{subfig}
\usepackage{textcomp}
\usepackage{stfloats}
\usepackage{url}
\usepackage{verbatim}
\usepackage{graphicx}
\usepackage{cite}
\def\BibTeX{{\rm B\kern-.05em{\sc i\kern-.025em b}\kern-.08em
    T\kern-.1667em\lower.7ex\hbox{E}\kern-.125emX}}
\usepackage{balance}
\usepackage{makecell}
\usepackage{multirow}
\usepackage{multicol}
\usepackage{color}
\usepackage[inline]{enumitem}
\usepackage{xcolor}

\makeatletter

\usepackage{fancyhdr}   
                            
\fancypagestyle{firstpage}{     
  \fancyhf{}
  \fancyfoot[C]{\small{\textcolor{gray}{~\copyright~ 2024 IEEE.  Personal use of this material is permitted.  Permission from IEEE must be obtained for all other uses, in any current or future media, including reprinting/republishing this material for advertising or promotional purposes, creating new collective works, for resale or redistribution to servers or lists, or reuse of any copyrighted component of this work in other works.}}}

}

\begin{document}
\title{On-Device Training Empowered Transfer Learning For Human Activity Recognition}
\author{Pixi Kang, Julian Moosmann, Sizhen Bian, and Michele Magno,~\IEEEmembership{Senior Member,~IEEE}
\thanks{Manuscript created October, 2020; This work was developed by the IEEE Publication Technology Department. This work is distributed under the \LaTeX \ Project Public License (LPPL) ( http://www.latex-project.org/ ) version 1.3. A copy of the LPPL, version 1.3, is included in the base \LaTeX \ documentation of all distributions of \LaTeX \ released 2003/12/01 or later. The opinions expressed here are entirely that of the author. No warranty is expressed or implied. User assumes all risk.}}

\markboth{Journal of \LaTeX\ Class Files,~Vol.~18, No.~9, September~2020}%
{How to Use the IEEEtran \LaTeX \ Templates}

\maketitle

\begin{abstract}
Human Activity Recognition (HAR) is an attractive topic to perceive human behavior and supplying assistive services. Besides the classical inertial unit and vision-based HAR methods, new sensing technologies, such as ultrasound and body-area electric fields, have emerged in HAR to enhance user experience and accommodate new application scenarios.
As those sensors are often paired with AI for HAR, they frequently encounter challenges due to limited training data compared to the more widely IMU or vision-based HAR solutions. Additionally, user-induced concept drift (UICD) is common in such HAR scenarios. UICD is characterized by deviations in the sample distribution of new users from that of the training participants, leading to deteriorated recognition performance.
This paper proposes an on-device transfer learning (ODTL) scheme tailored for energy- and resource-constrained IoT edge devices.
Optimized on-device training engines are developed for two representative MCU-level edge computing platforms: STM32F756ZG and GAP9. Based on this, we evaluated the ODTL benefits in three HAR scenarios: body capacitance-based gym activity recognition, QVAR- and ultrasonic-based hand gesture recognition. We demonstrated an improvement of 3.73\%, 17.38\%, and 3.70\% in the activity recognition accuracy, respectively.
Besides this, we observed that the RISC-V-based GAP9 achieves 20x and 280x less latency and power consumption than STM32F7 MCU during the ODTL deployment, demonstrating the advantages of employing the latest low-power parallel computing devices for edge tasks.

\end{abstract}


\input{c-intro}
\input{c-related_work}
\input{c-methodology}
\input{c-dataset}
\input{c-experiment}
\input{c-conclusion}

\bibliographystyle{IEEEtran}
\bibliography{references}{}

\begin{IEEEbiography}[{\includegraphics[width=1in,height=1.25in,clip,keepaspectratio]{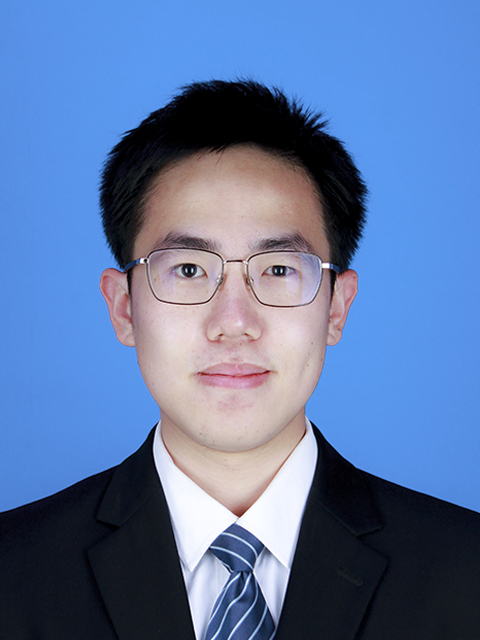}}]{Pixi Kang}
was born in Shandong, China in 1999. He received his B.E. degree in microelectronics science and engineering from Tsinghua University, Beijing, China in 2021, where he is currently pursuing his M.S. degree in integrated circuits engineering.
His research interests include smart sensing for human-centered applications, embedded systems and machine learning for edge devices.

\end{IEEEbiography}

\begin{IEEEbiography}[{\includegraphics[width=1in,height=1.25in,clip,keepaspectratio]{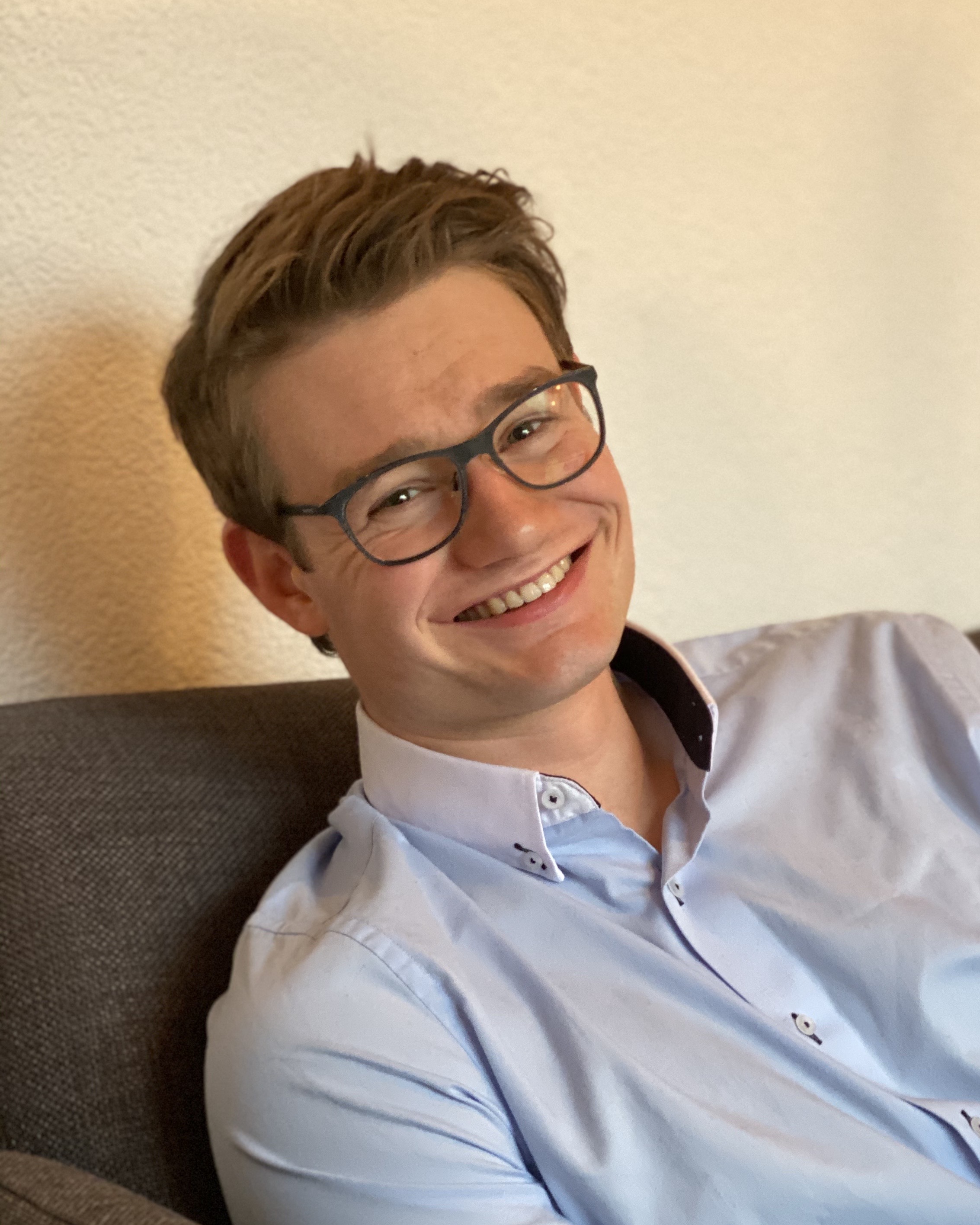}}]{Julian Moosmann} is a received the B.Sc. and M.Sc. degrees in electrical engineering and information technologies from ETH Zürich, Zürich, Switzerland, in 2019 and 2023, respectively. From 2022 to 2023, he was a Research Assistant at the Center for Project-Based Learning D-ITET, ETH Z\"urich, Z\"urich, Switzerland, where he is currently conducting his doctorate for pursuing the degree for Doctor of Science with the Integrated Systems Laboratory in conjunction with the Center for Project-Based Learning D-ITET. 

His research interests include a combination of computer vision, event-based sensing, low-power systems, wireless sensor networks, tiny machine learning / onboard intelligence, and battery-operated distributed systems.
\end{IEEEbiography}

\begin{IEEEbiography}
[{\includegraphics[width=1in,height=1.30in,clip]{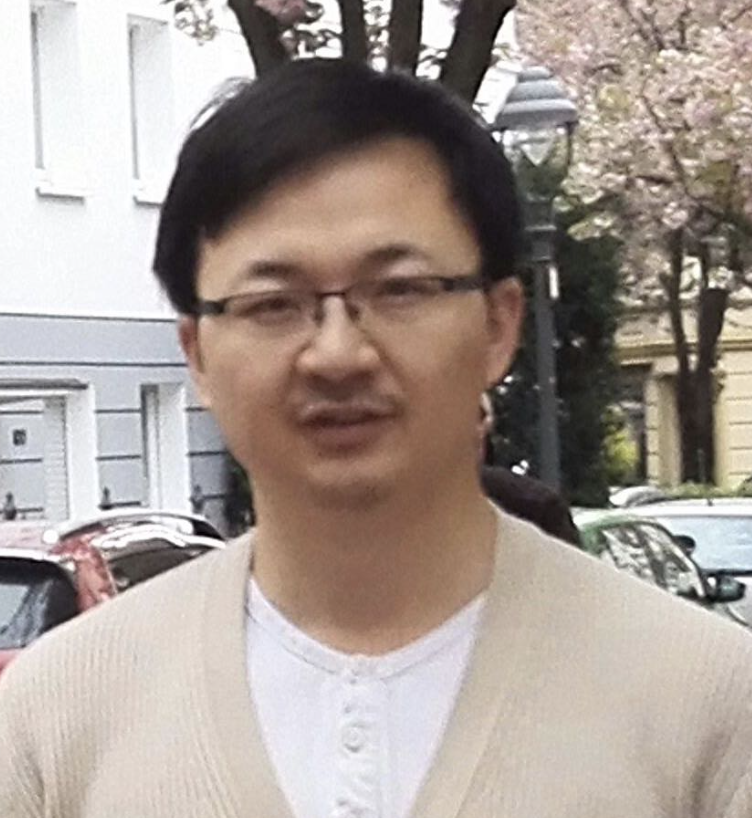}}]{Sizhen Bian} is a PostDoc at ETH Zurich in Switzerland since 2022. He received the Ph.D. degree in Computer Science at the University of Kaiserslautern-Landau and Embedded Intelligence in the German Research Center for Artificial Intelligence(DFKI), Germany in 2022, and the B.S. and M.S. degrees from the Northwestern Polytechnical University, China, and the University of Kaiserslautern-Landau
, Germany, in 2013 and 2016, respectively. His research interests are spiking neural networks and ubiquitous computing.
\end{IEEEbiography}

\begin{IEEEbiography}
[{\includegraphics[width=1in,height=1.25in,clip,keepaspectratio]{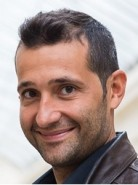}}]{Michele Magno}
is currently a Senior Scientist at ETH Zurich, Switzerland, at the Department of ¨ Information Technology and Electrical Engineering (D-ITET). Since 2020, he is leading the D-ITET center for project-based learning at ETH. He received his master’s and Ph.D. degrees in electronic engineering from the University of Bologna, Italy, in 2004 and 2010, respectively. He is working in ETH since 2013 and has become a visiting lecturer or professor at several universities, namely the University of Nice Sophia, France, Enssat Lannion, France, Univerisity of Bologna and Mid University Sweden, where currently is a full visiting professor at the electrical engineering department. His current research interests include smart sensing, low-power machine learning, wireless sensor networks, wearable devices, energy harvesting, low-power management techniques, and extension of the lifetime of batteries-operating devices. He has authored more than 220 papers in international journals and conferences. He is a senior IEEE member and an ACM member. Some of his publications were awarded as best papers awards at IEEE conferences. He also received awards for industrial projects or patents.
\end{IEEEbiography}

\end{document}

%% file: c-intro.tex
\section{Introduction}

\thispagestyle{firstpage}

Human activity recognition (HAR) is a technique that interprets user intentions or states through sensing and tagging user behavior patterns \cite{saleem2023toward, singh2023recent}. Integrated into Internet of Things (IoT) devices, HAR can serve as the core for a spectrum of applications like human-computer interaction\cite{wang_environment-independent_2021}, health surveillance\cite{bharathiraja_real-time_2023}, safety control\cite{dileep_suspicious_2022} and identity verification\cite{batool_authentication_2020}, attracting growing attention from both academia and industry.
As a branch of Cyber-Physical System (CPS)\cite{cassis_intelligent_2022}, HAR systems are characterized by sensors deployed at the system front end for data acquisition from various physical domains. 
A wide range of sensors have been proposed in the recent literature for HAR systems, e.g., optical camera\cite{shi2023robust}, IMU\cite{hao2023valerian}, sEMG\cite{tamulis2022affective} and radio-frequency sensors\cite{ding2023sparsity}.
The choice of sensing modality determines which aspects of the human motion should be captured and analyzed, hence directly influencing the recognition accuracy, user experience, application scenario, and cost of the system. 
Despite the dominance of vision- and IMU-based solutions, there is a growing demand for introducing new sensing modalities to HAR to address the limitation of vision and IMU-based solutions (accumulated error \cite{wei2021real}, line of sight \cite{ishioka2020single}, etc.) and enhance the sensing capability (range \cite{zhou20244d}, sensitivity \cite{yu2022noninvasive}, etc.).
For example, wireless sensors such as radar and ultrasonic\cite{abedi2023ai, ling_ultragesture_2022} have exhibited advantages such as reduced risk of privacy leakage and independence to light conditions when compared with optical-based solutions in realizing contact-free HAR. 
Besides that, researchers have proposed to use the body-area static electric field to extend the ability of solely IMU-based HAR while maintaining low cost and low power consumption \cite{reinschmidt2022realtime, bian2024body}. 
Innovations in sensing modalities are a major driving force, helping to push HAR's performance boundary further.

On the algorithm side, deep learning has gained unprecedented popularity and it is promising to be embedded into the HAR pipeline for sensor data analysis, benefitting from its impressive performance in feature extracting \cite{wang2023negative, yang2022efficientfi}.
On the other side, in applications targeting IoT devices, significant challenges arise \cite{mcenroe2022survey, wahab2022intrusion, kong2022edge}. These devices are typically battery-operated and constrained in terms of computational power and memory resources, making the deployment of resource-intensive neural network models particularly challenging \cite{ravaglia_tinyml_2021}. This limitation significantly impacts the feasibility of employing complex deep-learning models directly on edge devices. To bridge the gap between cutting-edge AI technologies and the capabilities of edge platforms, the field of Tiny Machine Learning (TinyML) has emerged as a promising solution \cite{ren_tinyol_2021}. Techniques such as quantization \cite{prakash2022iot}, network pruning \cite{lu2023snpf}, and the development of memory-efficient network architectures \cite{wang2023coopfl, choi2023simplification} have been developed to create compact models that are suitable for deployment on resource-limited devices.

In support of these advancements,  newly released microprocessors and microcontrollers (MCUs) have integrated various acceleration engines to promote the neural network and deep-learning model deployment, further bringing intelligence to IoT devices. Among others, MAX78000 \cite{MAX78000} from Analog Devices, equips an ultra-low-power CNN accelerator that has its memories for weights and biases and works as a peripheral to the main cores; NDP100 \cite{NDP100} from Syntiant, has a neural decision processor specifically designed to run deep learning models pursuing efficiency and throughput. Other examples are Hailo-8 \cite{Hailo8} from Hailo AI, STM32N6 \cite{STM32N6} from ST Semiconductor, Spresense from Sony, and many others. Among them, the GAP series from Greenwaves Technology, powered mainly by a cluster of low-power RISC-V cores, have been proven to be among the first-class for edge intelligent applications \cite{giordano2022survey, moosmann2023ultra, bian2022exploring}. The latest product, GAP9, utilizes a multi-core RISC-V cluster combined with a neural network acceleration called NE16, achieving a perfect balance between ultra-low power consumption, latency, flexibility, and ease of use. A chain of toolkits for neural network compiling, optimizing, and deploying is supplied for easier and more secure system-level application development. This work will be mainly evaluated on GAP9, aiming to pursue the edge computing efficiency regarding inference latency and energy consumption and benchmark the inference performance against a general-purpose MCU. 

This technological progression underscores the critical need for continued research and development in TinyML to fully leverage the capabilities of edge computing in HAR applications \cite{lattanzi2022exploring}. This approach not only addresses the challenges posed by limited device resources but also opens up new avenues for innovation in IoT-based applications \cite{incel2023device, saha2022machine}.
 
In many scenarios, such as personal wearable devices, smart homes, and smart cockpits, HAR systems are designed to serve single users, where a biased performance towards specific individuals is favorable \cite{craighero2023device, hayajneh2024tinyml}.
Nonetheless, due to the diversity of human behaviors, considerable variety can be observed in sensor data distributions across individuals, even against the same activity \cite{ferrari2020personalization}.
The data distribution in the user domain can easily deviate from that of the training set contributed by other individuals, causing the offline-trained deep learning model to perform degraded after deployment \cite{an2023transfer, an2023transfer}.
This issue can be referred to as User-Induced Concept Drift (UICD) \cite{mathur2020scaling}, which is dominant in HAR systems employing new sensors that have scarce data. 
Using the samples subsequently collected from the user domain after deployment to adjust the model subject to the user characteristics incrementally is a promising approach for improving robustness and generalization \cite{kwon2023lifelearner, chowdhary2023sensor}.  For example, In \cite{wang2020accurate}, the authors applied transfer learning on the Physionet EEG Motor Imagery dataset by first training and validating over the subjects via a five-fold cross-validation, getting a generalized model, and this model is then used as the backbone. The subject-specific training and validation via a four-fold inter-subject generated the final personalized models based on the backbone, achieving SOTA 65.7\% classification accuracy in a four-class task using the same technology. 

However, existing solutions for model deployment on MCU only support inference but lack the ability to perform post-deployment in-situ adaptation to the model. This means that user data needs to be uploaded to the server for processing and that the modified model should be re-downloaded to the edge device, resulting in privacy issues and extra power dissipation from data transmission\cite{ren_tinyol_2021}. On-device learning methods are emerging for low-end edge devices \cite{craighero_-device_2023, llisterri_gimenez_-device_2022}, their role in improving robustness and generalization in HAR tasks that utilize the newly explored sensing modalities has rarely been emphasized. On-device learning methods are crucial for advancing real-time, energy-efficient applications on low-end edge devices, enabling autonomous and responsive operations in environments with limited connectivity and computational resources and they can leverage low-power parallel edge computing processors \cite{nadalini2023reduced}, e.g., GAP9, to enache also latency and energy efficiency.

This paper introduces a novel on-device transfer learning (ODTL) approach to address the UICD problem in HAR systems, specifically tailored for IoT devices constrained by MCU-level computation and memory resources. This innovative method leverages transfer learning directly on the device, significantly enhancing the efficiency and accuracy of HAR tasks while operating within stringent hardware limitations. By pushing the boundaries of what is possible with MCU-level devices, our solution contributes a pioneering framework that paves the way for more advanced, autonomous IoT applications. The proposed solution is also suitable for other sensors data and application. The contributions of this work can be summarized as follows:
\begin{itemize}
    \item \textbf{UICD Impact in HAR:}  We quantitatively analyze the performance loss caused by UICD in three HAR scenarios: human body capacitance (HBC)+IMU-based gym activity recognition, electrostatic charge variation (QVAR)+IMU-based hand gesture recognition and 40kHz ultrasonic-based hand gesture recognition.
    \item \textbf{ODTL Engines at the Edge:} We proposed a quantized lightweight ODTL engines. The proposed model ahs been optimized and deployed on ARM Cortex-M cores ( STM32F7-series) and Risc-V cores, specifically on the novel parallel processors  GAP9 from GreenWaves Technology, making extensions to the software stacks of respective neural network model deployment toolkits.
    \item \textbf{Effectiveness of ODTL in Solving UICD:} We demonstrate the effectiveness of using sample stream from user domain for real-time ODTL at the edge in improving personalized recognition performance.
    \item \textbf{Latency and Power Efficiency Evaluation:} We carry out extensive experiments to evaluate the performance of inference and ODTL methods on the two platforms regarding latency and power efficiency. Our results show that the RISC-V-based GAP9 achieves 20x and 280x less latency and power consumption than STM32F7 MCU during the ODTL deployment, demonstrating the superiority of using the latest parallel architecture for edge inference. 
\end{itemize}

%% file: c-related_work.tex
\section{Related Works}

\subsection{Sensor-based human activity recognition}
The choice of sensing modality plays an important role in deciding the performance of a HAR system \cite{fu2020sensing, bian2022state}.
As an intuitive approach to recording human behavior, the application of image sensors, e.g., RGB\cite{gowda_human_2017} and RGB-D\cite{agarwal_weighted_2015} cameras, has been extensively studied for HAR tasks.
Such solutions benefit from a contact-free interaction logic during operation\cite{beddiar_vision-based_2020}.
Meanwhile, thanks to the rapid development in the field of computer vision and the standardized image data format, there are abundant resources of advanced DL models and publicly available image/video data for training.
Nonetheless, image sensor-based solutions suffer from privacy issues, which makes them less suitable for private scenarios. In addition to the image sensors, IMU is another dominant sensing modality used in HAR. It benefits from advantages such as miniature size, low cost, low power consumption, and ubiquity.
Despite the rich kinematic information (e.g., acceleration and angular velocity) IMU provides for modeling human activities, its sensing field is restricted to the device wearer,  while it stays blind to the environmental and interpersonal contexts potentially correlated with the activity's category\cite{suh_worker_2023}. Given the limitations of these sensor solutions, it is imperative to expand the range of available sensing modalities for HAR to improve system performance further.

In recent years, wireless sensors such as radar\cite{pearce2023multi, chioccarello2023forte}, Wi-Fi\cite{liu2024unifi, liu2024unifi} and ultrasonic\cite{lian2023echosensor,ling_ultragesture_2022} have emerged as a strong competitor to image sensors in realizing high-precision contact-free HAR.
Instead of images/videos, these methods collect the echoes of electromagnetic or acoustic waves against the human body, benefiting from the low interpretability of private behaviors and the insensitivity against illumination conditions\cite{hussain2023low}.
In particular, ultrasonic-based solutions are attracting growing attention due to their capabilities of achieving finer range resolution, simplifying the transceiver circuits, and staying insensitive to electromagnetic interference\cite{zhou_efficient_2020}.
For example, Ling \textit{et al.} propose UltraGesture\cite{ling_ultragesture_2022}, a hand gesture recognition system built upon commodity speakers and microphones on mobile devices. 
Using Channel Impulse Response (CIR) for data representation, UltraGesture achieves a range resolution of 7mm and recognizes 12 fine-grained hand gestures with an average accuracy of over 99\%.
In addition to hand gesture recognition, ultrasonic-based HAR systems have also been built for a variety of other applications such as in-air digits recognition\cite{yang2023ultradigit} and head tracking\cite{wang2022faceori}, exhibiting high potentials in HAR tasks featuring short-distance interaction and fine-grained movements. In the wearable domain, IMU is the dominant or even the sole sensor for motion sensing in current commercial products. However, IMU is mainly used for the motion pattern sensing of the body part to which it is being attached. 
Recently, researchers have explored an alternative wearable body motion sensing modality for full-body motion sensing without the limitation of sensor deployment place, the human body capacitance (HBC) \cite{bian2022contribution}. The corresponding sensor consumes less power than IMU and can be manufactured in an ultra-small form factor. 
HBC has two properties that make it an attractive complement to IMU. First, the deployment of the sensing node on the being tracked body part is not a requirement for the HBC sensing approach. Thus, for example, a wrist-worn HBC sensor can be used to track and recognize leg-based exercises \cite{bian2022using}. Furthermore, HBC is sensitive to the body's interaction with its surroundings, including both touching and being in the immediate proximity of people and objects \cite{bian2019wrist}. 
Based on the emerging exploration and impressive achievement of body-area capacitive sensing, the industry has designed and manufactured some novel chip-level capacitive sensors. A typical example is the Qvar sensor released by ST Semiconductor in 2021. Benefitting from its ultra-low power consumption and ultra-small form factor, some application works based on Qvar have been published \cite{reinschmidt2022realtime, dheman2022cardiac, manoni2022long}. Despite the impressiveness of such novel sensing modalities in HAR applications, the practical deployment often faces the UICD problem, and the lack of training data is an unneglectable issue when expecting a robust inference.

\subsection{Gradient descent-based on-device continuous learning}

\begin{table*}[htbp]
\centering
\caption{On-Device Training Methods for Edge Platforms}
\label{table:on_device_training_methods}
\resizebox{\textwidth}{!}
{
    \begin{tabular}{|c|c|c|c|c|c|c|c|}
    \hline
    \textbf{Proposed Method}                                               & \textbf{Sensing Modality} & \textbf{Target Task}              & \textbf{Target Device/Core}      & \textbf{SRAM} & \textbf{Flash} & \textbf{ML Algorithm} & \textbf{Full Training} \\
    \hline
    \multicolumn{8}{|c|}{\textbf{Non-Gradient Descent-Based}}\\
    \hline
    De Vita \textit{et al.}, $\mu$-FF\cite{de_vita_-ff_2023}               & Optical                   & Image classification              & Arm Cortex-M7@480MHz             & 1MB           & 2MB            & FF+FC                 & No \\
    Pau \textit{et al.}\cite{pau_automated_2022}                           & Motion MEMS               & Human activity recognition        & Arm Cortex-M7@480MHz             & N.A.          & N.A.           & 1D-CNN+RCE-NN         & No \\ 
    Disabato \textit{et al.}\cite{disabato_incremental_2020}               & Audio                     & Speech recognition                & Arm Cortex-M7@216MHz             & 512KB         & 2MB            & 2D-CNN+kNN            & No \\
    Pau \textit{et al.} \cite{pau_online_2021}                             & Environmental             & Anomaly detection                 & Arm Cortex-M7@400MHz             & 1MB           & 2MB            & DeepESN+FC            & No \\ 
    Abdennadher \textit{et al.}\cite{abdennadher_fixed_2021}               & Biopotential              & Anomaly detection                 & Arm Cortex-M7@480MHz             & N.A.          & N.A.           & RC+OC-SVM             & No \\ 
    \hline
    \multicolumn{8}{|c|}{\textbf{Gradient Descent-Based}}\\
    \hline
    Lin \textit{et al.}, TTE\cite{lin_-device_2022}                        & Optical                   & Image classification              & Arm Cortex-M7@216MHz             & 320KB         & 1MB            & 2D-CNN                & Yes  \\
    Khan \textit{et al.}, DaCapo\cite{khan_dacapo_2023}                    & Optical                   & Image classification              & Arm Cortex-M4@180MHz             & 256KB         & 2MB            & 2D-CNN                & Yes  \\
    Kwon \textit{et al.}, TinyTrain\cite{kwon_tinytrain_2023}              & Optical                   & Image classification              & Raspberry Pi Zero 2, Jetson Nano & -             & -              & 2D-CNN                & Yes  \\
    Ravaglia \textit{et al.}, QLR-CL\cite{ravaglia_tinyml_2021}            & Optical                   & Image classification              & VEGA\cite{rossi_44_2021}         & -             & -              & 2D-CNN                & No   \\
    Kopparapu \textit{et al.}, TinyFedTL\cite{kopparapu_tinyfedtl_2022}    & Optical                   & Image classification              & Arm Cortex-M4@\ 64MHz            & 256KB         & 1MB            & 2D-CNN                & No   \\
    Avi \textit{et al.}\cite{avi_incremental_2022}                         & IMU& Handwritten letter recognition & Arm Cortex-M4@\ 84MHz            & 96KB          & 512KB          & FCNN                  & No                                    \\
    Ren \textit{et al.}, TinyOL\cite{ren_tinyol_2021}                      & IMU& Anomaly detection                 & Arm Cortex-M4@\ 64MHz            & 256KB         & 1MB            & FCNN                  & No                                    \\
    Craighero \textit{et al.}\cite{craighero_-device_2023}                 & Motion MEMS               & Human activity recognition        & Arm Cortex-M4@\ 80MHz            & 320KB         & 1MB            & 1D-CNN                & Yes  \\ 
    Llisterri Giménez \textit{et al.}\cite{llisterri_gimenez_-device_2022} & Audio                     & Keyword spotting                  & Arm Cortex-M4@\ 64MHz            & 256KB         & 1MB            & FCNN                  & Yes  \\
    \hline
    \end{tabular}
}
\end{table*}
Given the diversity and instability of the deployment scenarios of IoT devices, it is frequently desirable to fine-tune or re-train the integrated ML algorithms locally to handle drifted data distributions while preserving user privacy.
Nonetheless, the highly restricted computation and memory resources of IoT devices, typically low-power MCUs with SRAMs of several kilobytes, have made this task challenging, and the number of works dedicated to this topic is quite limited.
Existing methods for on-device learning on MCUs can be divided into two categories according to whether the iterative gradient descent method is involved, as summarized in Table \ref{table:on_device_training_methods}.
Backpropagation (BP) is the canonical method for training a multi-layer neural network, where the model parameters are updated iteratively based on the gradients computed with respect to the defined loss.
Due to the high computation and memory overheads induced by BP, researchers have proposed surrogate solutions targeting MCU deployment.
\cite{pau_online_2021} and \cite{de_vita_-ff_2023}  provide a one-off learning process by using Normal Equation (NE) to find the closed-form solution for the multivariate Ridge regression problem at the last stage of the network.
\cite{pau_automated_2022} adopts Restricted Coulomb Energy Neural Network (RCE-NN) as the target model for training, which involves only the forward propagation of data, expediting the learning process.
Besides, k-Nearest Neighbors (k-NN)\cite{disabato_incremental_2020} and One-Class Support Vector Machine (OC-SVM)\cite{abdennadher_fixed_2021} have also exhibited the potential of realizing compact learning on MCUs.
Despite the lower requirement for resources, there is typically a gap in terms of generalization performance between these methods from the BP-based ones\cite{de_vita_-ff_2023}.
BP requires the storage of intermediate activations to compute the weight gradients, thus posing heavy, often unacceptable memory burdens to low-end edge devices.
Up till now, universal frameworks enabling full-network fine-tuning/training on MCUs\cite{lin_-device_2022, khan_dacapo_2023, kwon_tinytrain_2023} are still scarce, while they are mostly dedicated to vision tasks based on 2D-Convolutional Neural Network (CNN) architecture.
Compared with fine-tuning the whole network, on-device transfer learning (ODTL)\cite{ren_tinyol_2021, kopparapu_tinyfedtl_2022, avi_incremental_2022} that selectively updates only specific layers, e.g., the last dense layer, offers a preferable solution that better balances between resource-friendliness and performance gains.
For example, \cite{craighero_-device_2023} reports an average performance loss of only 1.5\% by shifting from full-network training to partial transfer learning (only dense layers) in the personalization task on the WISDM data set, while 70.4\% latency and 36.2\% memory footprint can be saved during the learning process.

Regarding the hardware platforms, general-purpose MCUs equipped with Arm Cortex-M cores are most frequently used for method verification.
The operating frequency of the cores ranges from 64MHz to 480MHz, while the available size of SRAM and FLASH is in the order of hundreds of KB and several MB, respectively.
In addition to the Arm-powered MCUs, some MCU-class multi-core edge processors have also been explored for similar tasks. An example is \cite{ravaglia_tinyml_2021}, where the authors implement an on-device continual learning framework on VEGA\cite{rossi_44_2021}, a PULP-based SoC combining parallel programming with ultra-low-power features.
Compared with an STM32 L4 MCU, VEGA proves to be 65$\times$ faster and 37$\times$ more energy efficient during learning.

%% file: c-methodology.tex
\section{Methodology}
The key components involved in the design and the deployment of the proposed ODTL are described as follows. First, the \textit{Network Topology} introduces the neural network's structure, which is pivotal for understanding the \textit{Rules of On-Device Parameter Update}---which details the processes and criteria for modifying the network parameters on-device. Last but not least, the \textit{On-device Training Engine for STM32F7 series MCU} is presented, followed by the \textit{On-device Training Engine for GAP9}. Both examine the training process of the two hardware architectures used.

\subsection{Network Topology}\label{network_topology}

\begin{figure}[htbp]
\centering 
\subfloat[]{\includegraphics[width=\columnwidth]{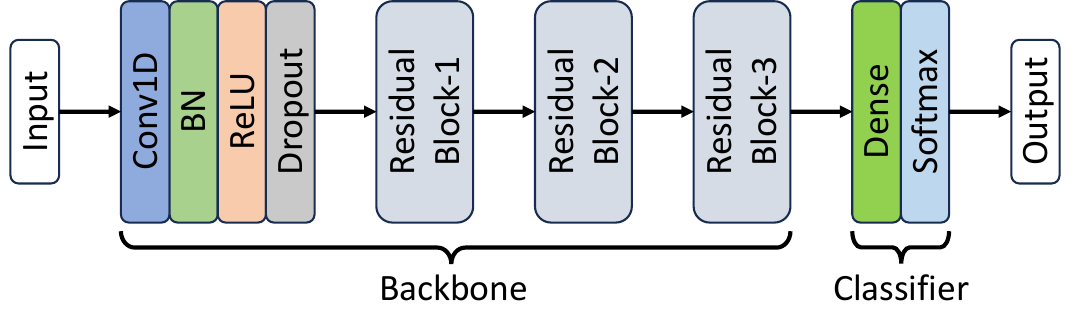}\label{fig:whole_network}}
\hfil
\subfloat[]{\includegraphics[width=\columnwidth]{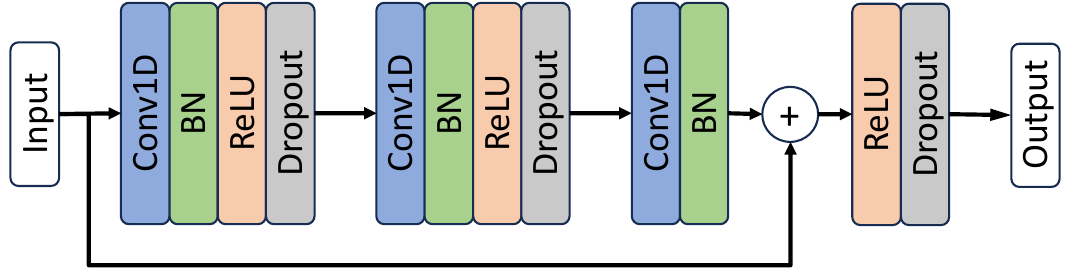}\label{fig:residual_block}}
\caption{Network topology. (a) Whole network. (b) Residual block.}
\label{fig:network_topology}
\end{figure}

Considering the format of the sensor recordings involved in this paper (multi-channel time series), a 1D-CNN model is adopted for classification.
Fig.\ref{fig:network_topology}(a) shows the employed model, which is a residual network consisting of 10 1D-convolutional layers and 1 fully connected (dense) layer\cite{bian2022exploring}.
The whole network can be divided into two parts: the backbone and the classifier.
The first layer of the backbone is a 1D convolutional layer, with the kernel size, stride, padding, and output channel number set as 3, 1, 1, and 32, respectively.
It maps input tensors of shape $(C_i, W_i)$ to tensors of shape $(32, W_i)$, where $C_i$ and $W_i$ represent the channel number and the width of the input data, respectively.
Following the first layer are Batch Normalization (BN), ReLU, and a dropout layer with rate of 0.1.
The rest of the backbone comprises three residual blocks, whose structure is illustrated in Fig.\ref{fig:network_topology}(b).
The configuration of the convolutional layers within the residual block is the same as that of the backbone's first layer, which means that the shape of the intermediate activations remains the same throughout the backbone.
The residual connection skips three convolutional layers and merges the block's input with the third BN's output through addition.
Then, there follows another pair of ReLU and dropout layers.
The output tensor of the backbone is of shape $(32, W_i)$, which is then flattened and fed to the classifier.
The classifier consists of a single dense layer and a softmax layer.
It maps feature vectors extracted by the backbone to probabilities over the defined classes.


\subsection{Rules of On-Device Parameter Update}
\label{subsec: rules_of_parameter_update}
After deployment of the pre-trained model onto MCUs, the parameters of the backbone are frozen and used only for inference purpose, while the classifier's parameters, including the weights and bias of the dense layer, are updated iteratively using the gradients calculated based on the input sample stream.
Each input sample is used once and discarded afterward, therefore getting rid of the need to store past samples.
Here, we introduce the rules for parameter update and its derivation.

Assume $x(\tau)$ is the feature vector extracted by the backbone at time step $\tau$, and the classifier maps $x(\tau)$ to a probability vector $P(\tau)$, which can be formulated as:
\begin{equation}
    P(\tau) = \text{softmax}(W(\tau)x(\tau)+b(\tau))
\end{equation}
where $W(\tau)$ and $b(\tau)$ are the weight matrix and the bias vector of the dense layer at time step $\tau$, respectively.
Specifically, the $i$th element of $P(\tau)$ can be calculated as:
\begin{equation}
    P(\tau, i)=\frac{e^{W(\tau, i)x(\tau)+b(\tau)}}{\sum_{k=1}^{C}{e^{W(\tau, k)x(\tau)+b(\tau)}}}
\end{equation}
where $C$ is the number of classes to be classified and $W(\tau,i)$ denotes the group of weights that connects with the $i$th probability output at time step $\tau$.
Therefore, given the label of the current input $y(\tau)$, the cross-entropy loss $L(\tau)$ can be calculated as:
\begin{equation}
    L(\tau) = \sum_{k=1}^{C}{-I(k=y(\tau))\log P(\tau, i)}
\end{equation}
where $I(\cdot)$ equals 1 if the condition within is met otherwise it equals 0.
The gradients needed for parameter update are then obtained by taking derivatives of the loss with respect to the corresponding parameters:
\begin{equation}
g_w(\tau,i,j)\leftarrow\frac{\partial L(\tau)}{\partial W(\tau,i,j)}=x(\tau,j)\left(P(\tau,i)-\mathbb{I}(y(\tau)=i)\right)
\end{equation}
\begin{equation}
g_b(\tau,i)\leftarrow\frac{\partial L(\tau)}{\partial b(\tau,i)}=P(\tau,i)-\mathbb{I}(y(\tau)=i)
\end{equation}
where $g_w(\tau,i,j)$ denotes the gradient of the weight that connects the $i$th element of the dense layer's output with the $j$th element of the layer's input, while $g_b(\tau,i)$ denotes the gradient of the bias that contributes to the $i$th output.

To expedite the learning process and to escape from local minima, we implement Stochastic Gradient Descent (SGD) with momentum as the optimizer to manage past gradients and perform parameter updates.
Concretely, the exponential moving averages (EMA) of the parameters' historical gradients are maintained and updated each time there inputs a new sample:
\begin{equation}
    i_w(\tau,i,j)\leftarrow \mu i_w(\tau-1,i,j)+g_w(\tau,i,j)
\end{equation}
\begin{equation}
    i_b(\tau,i)\leftarrow \mu i_b(\tau-1,i)+g_b(\tau,i)
\end{equation}
Where $i_w(\tau,i,j)$ and $i_b(\tau,i)$ denote the EMA of the corresponding weight and bias gradient up to $\tau$, respectively, while $\mu$ is the weight assigned to the past values of the gradients.
Then, the parameters of the dense layer are updated using:
\begin{equation}
    W(\tau+1,i,j)\leftarrow W(\tau,i,j)-\gamma i_w(\tau,i,j)
\end{equation}
\begin{equation}
    b(\tau+1,i)\leftarrow b(\tau,i)-\gamma i_b(\tau,i)
\end{equation}
where $\gamma$ is the learning rate.

\begin{figure*}[htbp]
\subfloat[]{
\begin{minipage}{0.475\linewidth}
    \centerline{\includegraphics[width=\textwidth]{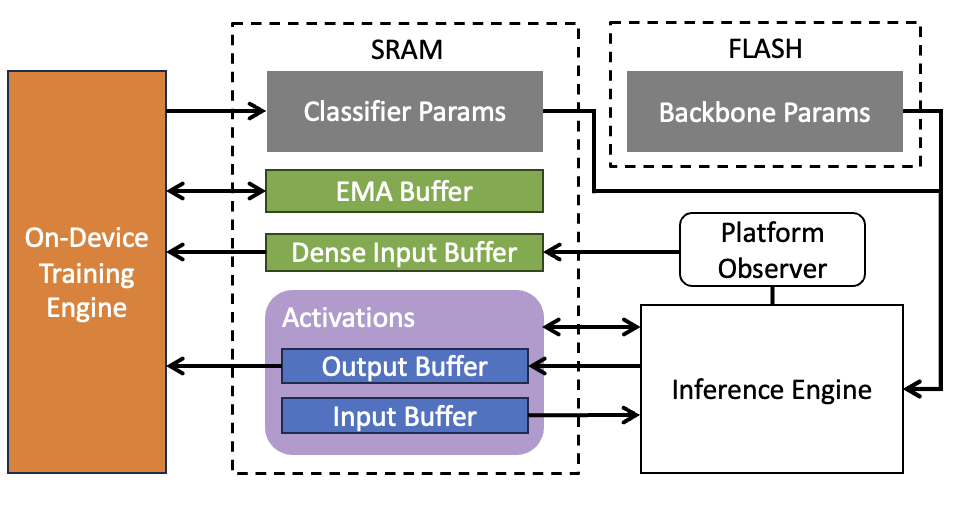}}
    \label{fig:ot-engine-stm32f7}
\end{minipage}
}
\subfloat[]{
\begin{minipage}{0.515\linewidth}
    \centerline{\includegraphics[width=\textwidth]{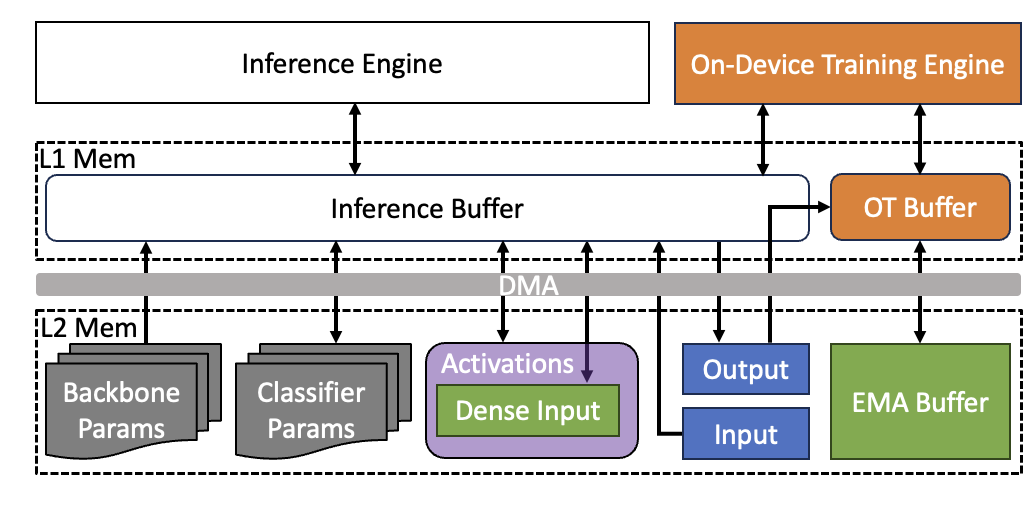}}
    \label{fig:otengine-gap9}
\end{minipage}
}

\caption{Architectures of implemented on-device training engines on STM32F7 series MCU and GAP9 processor. (a) STM32F7 MCU. (b) GAP9.}
\label{fig:ot-engines}
\end{figure*}

\subsection{On-device training engine for STM32F7 series MCU}
STM32F7 is a popular series of general-purpose MCUs from ST Microelectronics.
It features an Arm Cortex-M7 core operating up to 216MHz, which is embedded with FPU that supports floating point operations required by high-precision DL models.
Specifically, we employ a NUCLEO-F756ZG development board with an STM32F756ZG MCU, which has an internal SRAM memory of 320KiB and an internal Flash memory of 1MiB.
We consider this specification to be representative of the general-purpose MCUs that are equipped with AI processing capability, which guarantees the generalizability of the implemented on-device training method.

To import the pre-trained network to the MCU, X-CUBE-AI, a toolkit for automatic AI algorithms conversion for STM32 devices is used.
It takes as input pre-trained models formatted in ONNX and generates C-models that can be deployed onto the device.
It further provides a compiled runtime library and a set of APIs to realize efficient on-device inference.
Nonetheless, X-CUBE-AI does not support on-device training.
Accordingly, an on-device training engine is designed as an extension to the tool to realize post-deployment updates to the pre-trained network.

Fig.\ref{fig:ot-engines}(a) illustrates the architecture of the implemented on-device training engine for STM32F7 series MCU.
In order to avoid the extra delay and power dissipation induced by frequently writing to the Flash, the parameters of the model are split into two parts and stored separately at runtime: the parameters of the classifier are kept in the SRAM, while those of the backbone still reside in the Flash.
An activation buffer is allocated and used as a private heap by the network to store the intermediate results during inference. 
Moreover, two additional buffers (named as Dense Input Buffer and EMA Buffer in Fig.\ref{fig:ot-engines}(a)) are maintained in the SRAM at runtime to cache the input tensor to the dense layer and the EMA of the past gradients, respectively.

To enable the engine, a start-up function is implemented and forced to be called before the parameter update.
The function migrates the classifier's parameters to the SRAM and registers the new address. Besides, it registers a call-back function as the platform observer to retrieve the input tensor to the dense layer as the inference process meets the execution of the corresponding node.
After initialization, the engine is able to perform real-time updates to the classifier following each inference, using the rules specified in \ref{subsec: rules_of_parameter_update}.
Specifically, the parameters' gradients are first calculated using the Dense Input Buffer, Output Buffer, and the label, whose contents should have been updated with respect to the latest input sample. The EMA buffer is then updated using the obtained gradients. Finally, the engine updates the classifier's parameters using the values currently stored in the EMA Buffer.
As an optional operation, the implemented on-device training method can be attached as an accessory to the inference method and can be enabled/disabled whenever needed without interfering with the latter.

\subsection{On-device training engine for GAP9}
GAP9 is an edge processor from GreenWaves Technology designed for intelligent IoT applications.
Exploiting the parallel nature of DL algorithms, it uses multiple RISC-V cores combined with a dedicated acceleration hardware engine to empower DL-centered applications at the edge while maintaining ultra-low-power features.
Structurally, GAP9 is composed of an MCU-type controller (Fabric Controller, FC) and a compute cluster.
The FC part contains a 1.6MiB L2 SRAM memory, a piece of 2MiB non-volatile memory, and a single RISC-V core in charge of coordinating the activities on GAP9.
On the cluster side, nine identical RISC-V cores embedded with FPU are deployed, which share an L1 Tightly Coupled Data Memory (TCDM) of size 128KiB.
The execution of DL algorithms is delegated to the cluster, where tasks are forked to the cores and joined after completion.
During this process, the cores access data directly and only from the L1 memory, while the data movements between L1 and L2 memory are explicitly planned at compile time and realized using Direct Memory Access (DMA) units.

Similar to STM32's X-CUBE-AI, GreenWaves Technology provides a set of toolkits to facilitate DL model deployment, as well as kernel libraries and APIs to realize efficient on-device inference.
Specifically, NNTOOL takes as input ONNX models and generates optimized C-code that executes the model using the GAP kernel library, while AutoTiler generates the explicit arrangement for the parameter placement and movement across the memory hierarchy at runtime.
Likewise, GAP SDK optimizes solely for inference while lacking support for on-device model updates, which has driven us to design a dedicated on-device training engine for GAP9 that manipulates its parallelization capability.

The implemented architecture is shown in Fig.\ref{fig:ot-engines}(b).
During the program start-up, buffers for network parameters, activation, input/output, and EMA are allocated in L2 memory and initialized.
Due to the hierarchical memory structure of GAP9, data residing in lower-level memories should be first transferred to the L1 memory before they can be processed by the cluster cores.
For the natively supported inference task, data transfer patterns and memory structure are automatically generated by AutoTiler.
While for the on-device training task, the data flow and memory structure involved are manually designated.
Moreover, to benefit from the parallel computing architecture, a kernel that supports multi-thread parameter updates is implemented as an extension to the GAP neural network library and used for on-device training.
Once the training process is triggered, the related resources, including the classifier's parameters, the dense layer's input tensor, the output tensor, and the EMA, are transferred to the L1 memory via DMA, accessed/updated in parallel by the cluster cores and then written back to the L2 memory.
Besides, considering the limited size of L1 memory, the resources can be configured to be transferred and processed in tiles so as to balance between latency and memory consumption.

%% file: c-dataset.tex
\section{Sensor-Based Human Activity Data Sets}
The utilized datasets for investigating the effect of ODTL in mitigating UICD in NS-HAR are RecGym dataset\cite{RecGym}, QVAR-based hand gesture dataset\cite{reinschmidt2022realtime} and ultrasonic-based hand gesture dataset\cite{kangpx_40khz-ultrasonicdhgr-onlinesemi_2023}, which are described subsequently in the following subsections. 

\begin{figure*}[htbp]
\subfloat[]{
\begin{minipage}{0.33\linewidth}
    \centerline{\includegraphics[width=\textwidth]{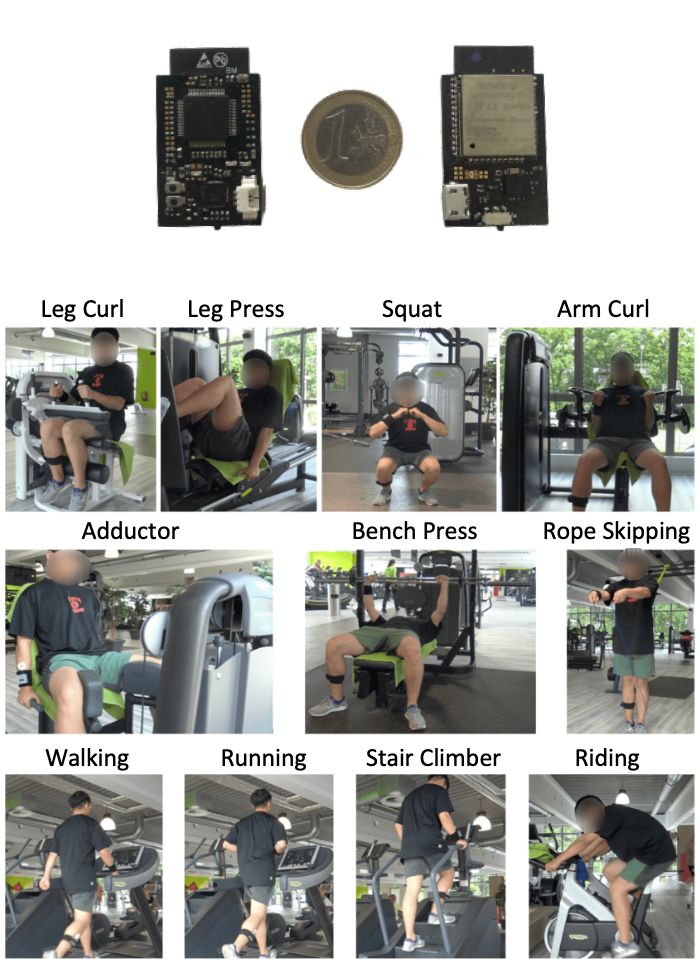}}
    \label{fig:gym}
\end{minipage}
}
\subfloat[]{
\begin{minipage}{0.33\linewidth}
    \centerline{\includegraphics[width=\textwidth]{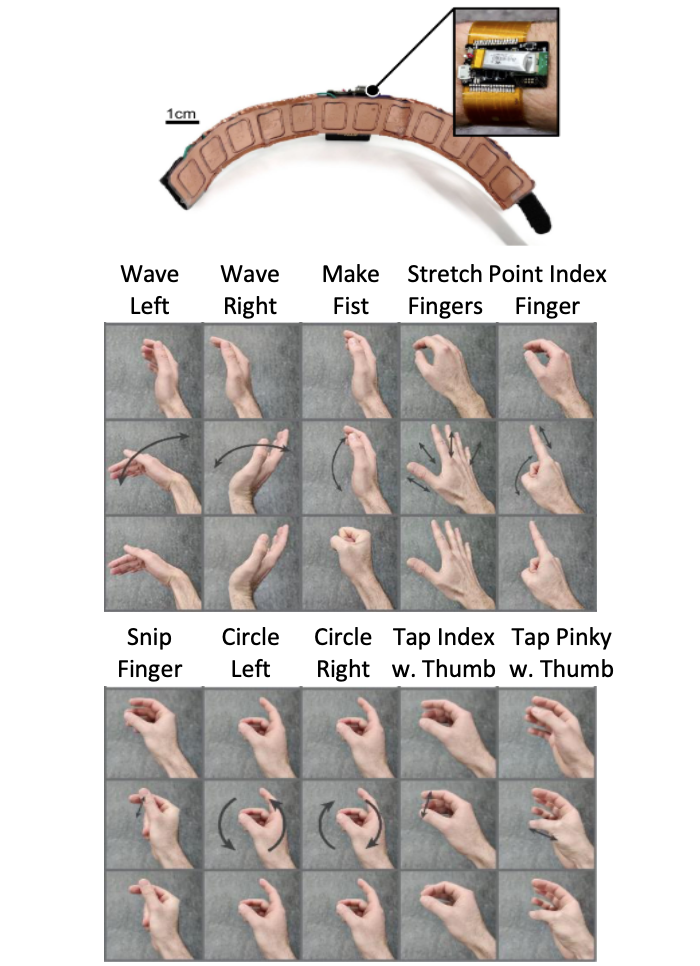}}
    \label{fig:qvar}
\end{minipage}
}
\subfloat[]{
\begin{minipage}{0.33\linewidth}
    \centerline{\includegraphics[width=\textwidth]{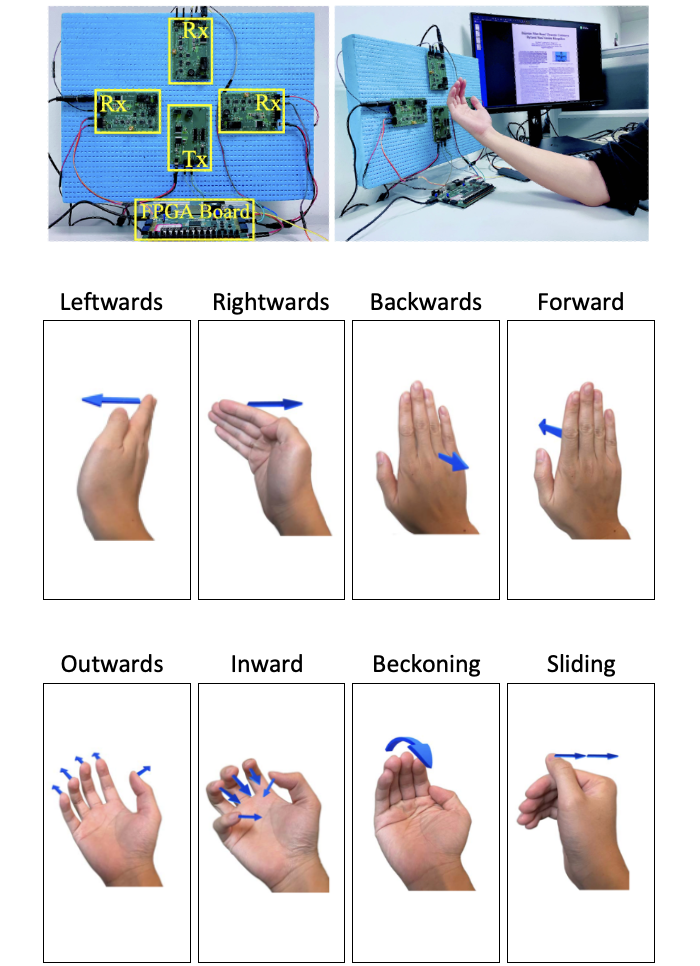}}
    \label{fig:ultra}
\end{minipage}
}
\caption{Novel sensor-based human activity data sets: devices and defined activities. (a) HBC+IMU-based gym activity recognition. (b) QVAR+IMU-based hand gesture recognition. (c) 40KHz ultrasonic-based hand gesture recognition.}
\end{figure*}

\subsection{RecGym: gym activity recognition}
RecGym\cite{RecGym} is a publicly available labeled data set, recording gym activities with sensing units composed of the inertial measurement unit and body capacitive sensor \cite{bian2019passive}. The sensing units were worn at three positions: on the wrist, in the pocket, and on the calf, sampling the motion context with a frequency of 20$Hz$. The data set contains the motion signals of twelve activities, including eleven workouts: adductor, arm curl, bench press, leg curl, leg press, riding, rope skipping, running, squat, stairs climber, walking, and a "null" activity when the volunteer hangs around between different workouts session. Each participant performed the above-listed workouts for five sessions in five days; each session lasted around one hour. Altogether, fifty sessions of gym workout data from ten volunteers are presented. A detailed description of the data set can be found in \cite{bian2022contribution}, in which the authors deployed a Resnet neural network for workout recognition and achieved an accuracy of 91\%. The authors also tried a real-time edge evaluation with the compressed network in \cite{bian2022exploring} and reported an accuracy of 88\%.

\subsection{QVAR: hand gesture recognition}
The QVAR dataset contains recordings of ten gestures from twenty individual participants, where every gesture was performed thirty times by each participant \cite{reinschmidt2022realtime}. The ten selected gestures are wave left, wave right, make a fist, stretch fingers, point index finger, snip fingers, circle left, circle right, tap index with the thumb, and tap pinky with the thumb. QVAR is a novel sensor that monitors the charge variation, released by ST Semiconductor in 2021. The principle behind this is to utilize the electrode attached to a high input impedance instrumentation amplifier for charge flow sensing. By placing an electrode on the wrist, any change in the distance between the electrode and the skin will result in electrostatic induction. Therefore, the sensor signal correlates with the movement pattern of the hand. Besides QVAR, the data set also includes the inertial measurement unit signal to enhance the recognition. Data was sampled with 240 $Hz$. With a three-dense layers neuron network, \cite{reinschmidt2022realtime} reports a gesture recognition of 87\% with cross-validation.

\subsection{Ultra: hand gesture recognition}
Ultra dataset\cite{kangpx_40khz-ultrasonicdhgr-onlinesemi_2023} is a publicly available data set disclosed in 2022, which contains a total of 5600 recordings of 8 predefined dynamic hand gestures, namely leftwards, rightwards, backwards, forward, outwards, inward, beckoning and sliding. The data acquisition system employs one ultrasonic transmitter, which emits coherent pulse trains with 40$kHz$  carrier frequency and 1.5$ms$ pulse repetition interval towards moving hand targets during operation. Three channels of echo signals are received and processed into Range-Doppler maps (RDMs) stream of frame rate 19$Hz$. A total of 45 frame-level features, e.g., maximum reflectivity and total instantaneous energy, are then extracted from each RDM. Seven volunteers participated in the data collection, during which they were asked to demonstrate each predefined gesture 100 times. Moreover, each demonstration should be completed within one second. Employing an OvR-random forest classifier with a feature alignment prefix, Zhou \textit{et al.}\cite{zhou_efficient_2020} reports a Leave-One-Person-Out cross-validation accuracy of 93.9\% on this data set.

%% file: c-experiment.tex
\section{Experiments and Results}
\subsection{Effect of UICD: A Quantitative Analysis}
\label{subsec:effect_of_uicd}
The effect of UICD manifests as the model's degraded recognition performance in processing inputs from users whose data are unavailable during offline training.
To quantize the loss caused by UICD, we use the model's recognition performance in the virtual case where the user samples can be accessed for offline training as the baseline while defining the performance gap between the two cases as the UICD loss.
Based on this metric, we first evaluate the UICD effects in the three NS-HAR scenarios.

Two different strategies are used to split the datasets to obtain the model's recognition performances under different user data accessibility during offline training:
\begin{itemize}
    \item {\it{Leave-One-Session-Out (L1SO)}}: Samples of the dataset are shuffled and split into $M$ equal-size sub-collections. $M$ rounds of model training/testing are carried out: in each round, one sub-collection is selected as the testing set, while the union of the rest sub-collections is used as the training/validation set.
    
    \item {\it{Leave-One-Person-Out (L1PO)}}: $N$ rounds of model training/testing are carried out: in each round, one individual is selected as the user, while the rest individuals are called non-users; all user samples compose the testing set, while all non-user samples compose the training/validation set.
\end{itemize}
To guarantee the fairness of the comparison, $M$ is set as the same value as $N$, i.e., the number of human contributors during data collection, for each dataset.
Based on the above dataset splitting strategies, the UICD loss is formulated as the difference between the averaged accuracy over all rounds of L1SO and the averaged accuracy over all rounds of L1PO:

\begin{equation}
    \mathcal{L}_{\text{UICD}} = \overline{Acc}_{\text{L1SO}} - \overline{Acc}_{\text{L1PO}}
\end{equation}


In each round of evaluation, the network introduced in \ref{network_topology} is trained with Adam optimizer with 0.001 learning rate, 0.9 $\beta_1$ and 0.999 $\beta_2$, combined with an early-stopping regularization with a patience value of 100.
Following the method introduced in \cite{bian2022exploring}, samples in the RecGym dataset are assigned with weighted importance due to the imbalanced label distribution. 
While in QVAR and Ultra datasets, all samples are considered to have equal importance during offline training.

The network's recognition performances on the three datasets are summarized in Table \ref{tab:offline_performance}.
Relatively higher performances are achieved on the three datasets under L1SO splitting, whereas different degrees of performance degradation, i.e., the UICD loss, can be observed in the corresponding L1PO tests.
Specifically, the UICD losses for RecGym, QVAR and Ultra are 0.74\%, 25.89\% and 6.00\%, respectively.
Among the three NS-HAR scenarios, gym activity recognition suffers the least from UICD, showing greater stability in dealing with previously unseen users.
On the one hand, this may due to the fact that the interaction with the gym equipment reduces the freedom of movements, thus resulting in lower varieties in activity characteristics across different individuals. 
On the other hand, it may also be due to the larger number of available training samples (around 1000$\sim$3000 per non-null class) in each round, which further inhibits the model from overfitting to the characteristics specific to non-users.
In contrast, tasks of QVAR and ultrasonic-based hand gesture recognition demonstrate a higher degree of vulnerability when confronted with unseen users, which can seriously affect the user experience, thus limiting the usability of related applications in real-world scenarios.

\begin{figure}[tb!]
    \centering
    \includegraphics[width=\columnwidth]{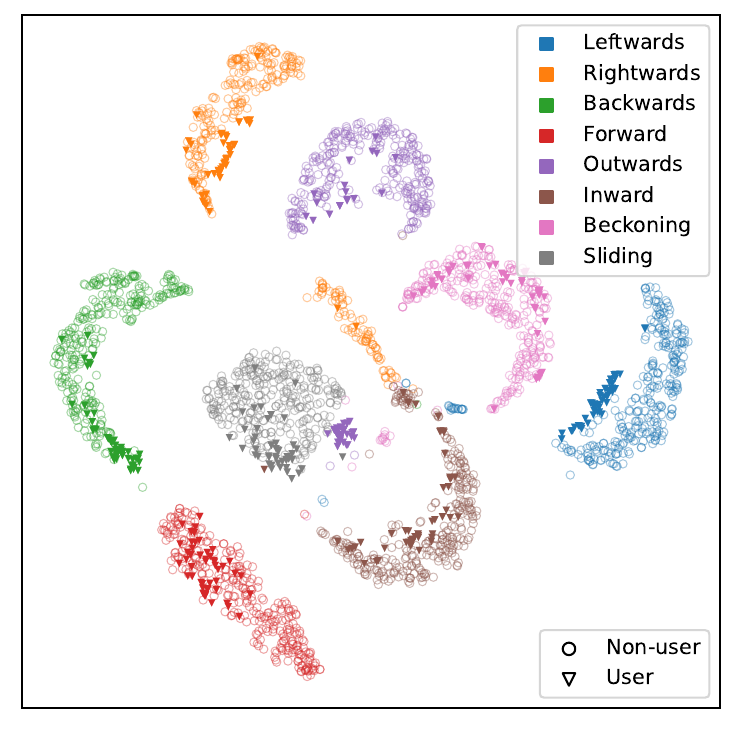}
    \caption{2-D t-SNE plot of the obtained embeddings in one round of L1PO test with individuals and classes distinguished by markers and colors, respectively.}
    \label{fig:ultra_tsne}
\end{figure}

In order to account for the UICD loss in an intuitive way, a 2-D t-SNE plot is shown in Fig. \ref{fig:ultra_tsne} to visualize the obtained embeddings in one round of the L1PO test on the Ultra dataset.
It can be observed that user samples of the same class tend to form clusters with relatively low inner variances, indicating high intra-individual stability.
Nonetheless, clusters representing the user's personalized characteristics may significantly deviate from the non-user samples of the same class, e.g., the lower cluster of "Outwards" is mapped closer to the cluster of "Sliding" than to its own class in Fig. \ref{fig:ultra_tsne}.
Consequently, samples from such clusters are prone to be misclassified by models trained upon non-user samples, which contributes to the UICD loss.
For HAR, the available data for offline training are usually limited in terms of individual diversity due to the high cost of data collection, which further increases the UICD risk.

\begin{table}[tb!]
    \centering
    \caption{Offline Training Performances under Two Different Data Splitting Strategies}
    \label{tab:offline_performance}
    \setlength{\tabcolsep}{3.5mm}
    \renewcommand\arraystretch{1.1}
    \begin{tabular}{|c|c|c|c|}
    \hline
    \rule{0pt}{9pt}
    Dataset     &  $\overline{Acc}_{\text{L1SO}}$  & $\overline{Acc}_{\text{L1PO}}$ & $\mathcal{L}_{\text{UICD}}$\\
    \hline
    RecGym      &  \textbf{91.04\%} & 90.30\% & 0.74\% \\
    QVAR        &  \textbf{92.34\%} & 66.45\% & 25.89\% \\
    Ultra       &  \textbf{98.29\%} & 92.29\% & 6.00\%\\
    \hline

    \end{tabular}
\end{table}

\subsection{ODTL for Mitigating UICD Effect}
In order to assess the capability of post-deployment ODTL in mitigating UICD effect, the datasets are further split based on the L1PO strategy:
\begin{itemize}
    \item \textit{L1PO-2}: $N$ rounds of model training/testing are carried out: in each round, one individual is selected as the user, while the rest of the individuals are called non-users; user samples are split into the post-deployment training set and testing set, while all non-user samples compose the offline training/validation set.
\end{itemize}
In practice, 40\% of the user samples are used for post-deployment training, while the remaining 60\% comprise the testing set.


In each evaluation round, the network is first trained with the offline training set on a GPU server, adhering to the specifications outlined in \ref{subsec:effect_of_uicd}.
For convenience, we refer to this stage of training as the Offline Training (OFT) in the rest part of paper.
After completing OFT, the trained models are then deployed onto the target hardware, i.e., STM32F756ZG and GAP9, for the subsequent ODTL process.
In order to meet the high precision demands during training, the deployed models are formatted in floating-point representation: float32 and bfloat16 on STM32F756ZG and GAP9, respectively.
When ODTL is initiated, the parameters of the network's classifier are updated using the sequentially incoming samples from the post-deployment training dataset following the rules specified in \ref{subsec: rules_of_parameter_update}.
Specifically, the learning rate and momentum used for RecGym, QVAR and Ultra are (0.002, 0.9), (0.002, 0.5) and (0.002, 0.5), respectively.
Taking into account the limited size of QVAR's post-deployment training set (around 10+ samples per class), the ODTL for QVAR extends over 5 epochs, while RecGym and Ultra each undergo 1 epoch of ODTL.

\begin{table}[tb!]
    \centering
    \caption{Performance Evaluation on STM32F756ZG and GAP9 after Offline Training and Post-Deployment Training (RecGym)}
    \label{tab:perf_eval_gym}
    \renewcommand\arraystretch{1.1}
    \begin{tabular}{|c|c|c|c|c|}
    \hline
    \multirow{2}*{Round} & \multicolumn{2}{c|}{STM32F756ZG} & \multicolumn{2}{c|}{GAP9} \\
    \cline{2-5}
    ~ & $Acc_{\text{OFT}}$ & $Acc_{\text{OFT+ODTL}}$ & $Acc_{\text{OFT}}$ & $Acc_{\text{OFT+ODTL}}$ \\
    \hline
    1 & 89.67\%& \textbf{94.34\%}& 89.60\%& \textbf{94.04\%}\\
    2 & 91.30\%& \textbf{93.82\%}& 91.30\%& \textbf{93.73\%}\\
    3 & 93.16\%& \textbf{95.38\%}& 93.39\%& \textbf{95.38\%}\\
    4 & 86.46\%& \textbf{90.73\%}& 86.80\%& \textbf{90.41\%}\\
    5 & 91.09\%& \textbf{94.99\%}& 91.11\%& \textbf{94.68\%}\\
    6 & 90.01\%& \textbf{92.95\%}& 90.09\%& \textbf{92.63\%}\\
    7 & 88.14\%& \textbf{93.53\%}& 88.14\%& \textbf{92.89\%}\\
    8 & 90.81\%& \textbf{94.90\%}& 91.08\%& \textbf{94.81\%}\\
    9 & 90.17\%& \textbf{93.86\%}& 90.15\%& \textbf{93.35\%}\\
    10 & 92.28\%& \textbf{95.89\%}& 92.28\%& \textbf{95.62\%}\\
    \hline
    Mean & 90.31\%& \textbf{94.04\%} &90.37\% & \textbf{93.75\%}\\
    \hline
    \end{tabular}
\end{table}

\begin{table}[tb!]
    \centering
    \caption{Performance Evaluation on STM32F756ZG and GAP9 after Offline Training and Post-Deployment Training (QVAR)}
    \label{tab:perf_eval_qvar}
    \renewcommand\arraystretch{1.1}
    \begin{tabular}{|c|c|c|c|c|}
    \hline
    \multirow{2}*{Round} & \multicolumn{2}{c|}{STM32F756ZG} & \multicolumn{2}{c|}{GAP9} \\
    \cline{2-5}
    ~ & $Acc_{\text{OFT}}$ & $Acc_{\text{OFT+ODTL}}$ & $Acc_{\text{OFT}}$ & $Acc_{\text{OFT+ODTL}}$ \\
    \hline
    1 & 72.58\%& \textbf{95.70\%} & 72.04\%& \textbf{95.70\%}\\
    2 & 84.57\%& \textbf{89.71\%}& 84.57\%&\textbf{88.57\%}\\
    3 & 51.12\%& \textbf{83.15\%}& 51.12\%&\textbf{82.58\%}\\
    4 & 36.13\%& \textbf{64.52\%}& 34.84\%&\textbf{63.23\%}\\
    5 & 53.45\%& \textbf{77.01\%}& 52.87\%&\textbf{78.16\%}\\
    6 & 60.00\%& \textbf{87.22\%}& 60.00\%&\textbf{86.67\%}\\
    7 & 59.20\%& \textbf{79.31\%}& 59.20\%&\textbf{79.31\%}\\
    8 & 93.89\%& \textbf{97.22\%}& 93.33\%&\textbf{97.78\%}\\
    9 & 85.45\%& \textbf{95.76\%}& 86.06\%&\textbf{96.36\%}\\
    10 & 46.15\%& \textbf{95.86\%}& 46.15\%&\textbf{94.67\%}\\
    11 & 89.33\%& \textbf{93.26\%}& 89.33\%&\textbf{92.70\%}\\
    12 & 85.00\%& \textbf{89.44\%}& 83.89\%&\textbf{89.44\%}\\
    13 & 85.80\%& \textbf{88.07\%}& 85.23\%&\textbf{86.93\%}\\
    14 & 80.57\%& \textbf{92.00\%}& 81.14\%&\textbf{91.43\%}\\
    15 & 89.89\%& \textbf{96.07\%}& 90.45\%&\textbf{96.63\%}\\
    16 & 56.50\%& \textbf{76.84\%}& 58.19\%&\textbf{74.01\%}\\
    17 & 45.45\%& \textbf{60.61\%}& 46.21\%&\textbf{59.85\%}\\
    18 & 34.83\%& \textbf{76.40\%}& 35.39\%&\textbf{74.16\%}\\
    19 & 59.20\%& \textbf{64.37\%}& 59.20\%&\textbf{64.37\%}\\
    20 & 73.30\%& \textbf{87.50\%}& 73.30\%&\textbf{87.50\%}\\
    \hline
    Mean& 67.12\%& \textbf{84.50\%}& 67.13\%& \textbf{84.00\%} \\
    \hline
    \end{tabular}
\end{table}

\begin{table}[tb!]
    \centering
    \caption{Performance Evaluation on STM32F756ZG and GAP9 after Offline Training and Post-Deployment Training (Ultra)}
    \label{tab:perf_eval_ultra}
    \renewcommand\arraystretch{1.1}
    \begin{tabular}{|c|c|c|c|c|}
    \hline
    \multirow{2}*{Round} & \multicolumn{2}{c|}{STM32F756ZG} & \multicolumn{2}{c|}{GAP9} \\
    \cline{2-5}
    ~ & $Acc_{\text{OFT}}$ & $Acc_{\text{OFT+ODTL}}$ & $Acc_{\text{OFT}}$ & $Acc_{\text{OFT+ODTL}}$ \\
    \hline
    1 &96.04\% & \textbf{98.96\%}&96.25\% &\textbf{98.75\%}\\
    2 & 89.58\%& \textbf{94.79\%}&89.38\% &\textbf{95.00\%}\\
    3 & 91.04\%& \textbf{93.96\%}&91.04\% &\textbf{93.96\%}\\
    4 & 87.92\%& \textbf{96.04\%}&88.13\% &\textbf{96.04\%}\\
    5 & 91.04\%& \textbf{93.96\%}&91.46\% &\textbf{94.38\%}\\
    6 & 89.79\%& \textbf{94.79\%}&89.58\% &\textbf{94.79\%}\\
    7 & \textbf{98.13\%}& 96.88\%&\textbf{98.13\%} &97.50\%\\
    \hline
    Mean & 91.93\%& \textbf{95.63\%}& 91.93\%& \textbf{95.69\%}\\
    \hline

    \end{tabular}
\end{table}

Table \ref{tab:perf_eval_gym}$\sim$\ref{tab:perf_eval_ultra} present the recognition performances achieved by the networks across the three datasets, which compares the results after completing only OFT and after completing both OFT and ODTL.
Notably, the network's recognition performance is generally improved when ODTL is included.
Taking the results obtained on STM32F756ZG as an example, RecGym, QVAR, and Ultra show an average accuracy improvement of 3.73\%, 17.38\%, and 3.70\%, respectively.
To validate the correctness of the implemented on-device training engines, we use a full-precision network trained on a GPU server under the PyTorch framework as the reference model and compare its performances with those obtained on the target devices.
Results show that the reference model's performances align precisely with those of its counterpart on STM32F756ZG.
Furthermore, despite using a truncated version of the floating-point format, no significant performance compromise is observed for the networks deployed on GAP9.

Among the three datasets, QVAR achieves the highest performance improvement when ODTL is included, indicating that this sensor facilitates a higher degree of personalization in realizing HAR.
It also indicates that the samples in QVAR exhibit low intra-individual variance, allowing the network to rapidly adapt to the user's activity characteristics based on representative data samples.
An interesting observation is that the gym activity recognition network outperforms its L1SO testing counterpart after completing ODTL.
We attribute this to the expanded amount of data used for training by incorporating additional samples from the post-deployment training set.
Moreover, the network's tendency to misclassify low-confidence "Null" samples into other classes, established during offline training due to the weighted importance assigning strategy, is rectified to some extent by forcing the classifier to be updated on a single input basis each time during ODTL, resulting in a further increase in accuracy.
While ODTL leads to performance improvement in the majority of testing rounds across all three datasets, the seventh round of testing of Ultra highlights a potential risk of performance degradation associated with ODTL.
Methods of mitigating such risk is out of the scope of this paper, while it may merit further discussion in future works.


\subsection{On-Device Latency and Power Evaluation}
The evaluation is performed for all the three datasets and networks with respect to the
\begin{enumerate*}[label=(\roman*),,font=\itshape]
\item the inference of the networks
\item and the parameter update of the networks.
\end{enumerate*}

\begin{figure}[h!]
    \centering
    \includegraphics[width=\linewidth]{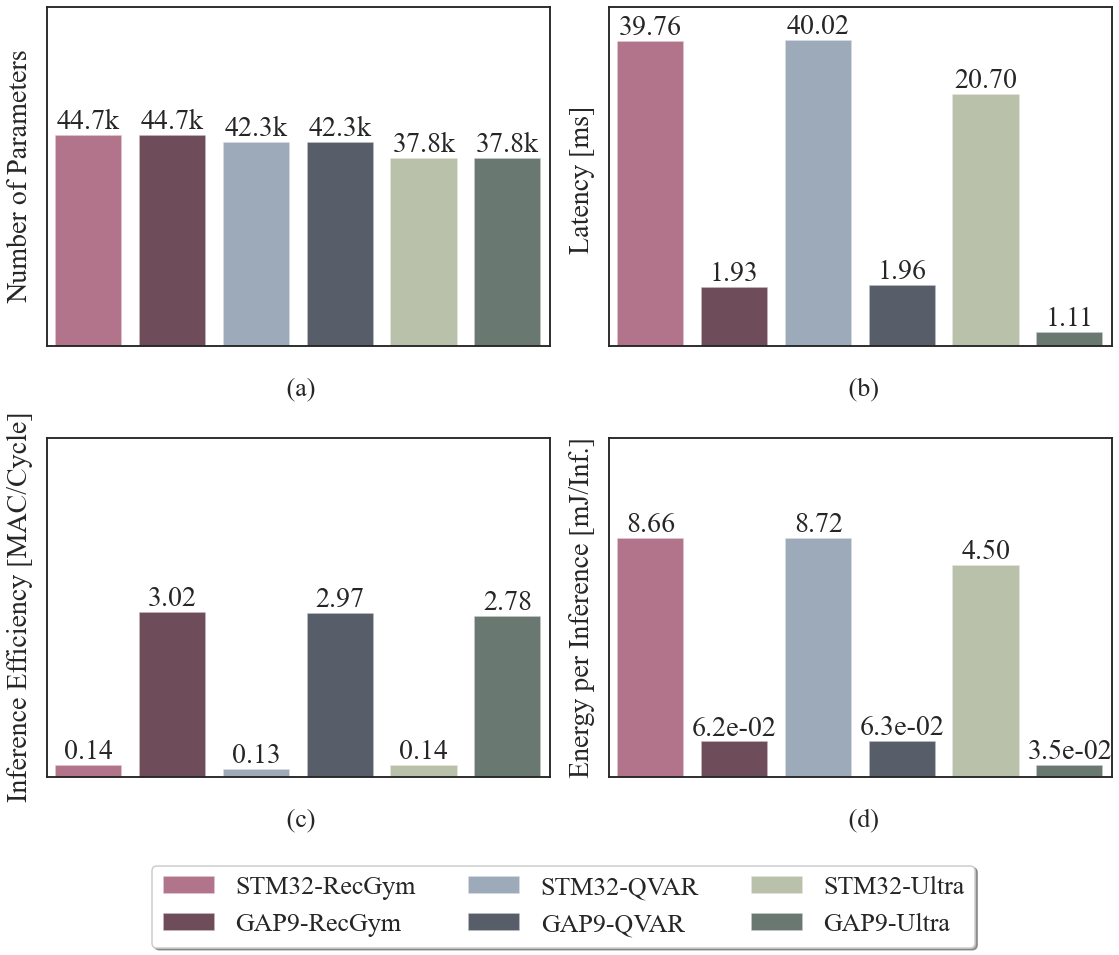}
    \caption{Comparison of the networks' on-device inference execution for all three datasets Gym, Qvar and Ultra. for the three datasets RecGym, Qvar and Ultra. The networks deployed on GAP9 are run with 200MHz while the networks deployed on STM32F756ZG are run with 216MHz.}
    \label{fig:ot_inference}
\end{figure}

\begin{figure*}[h!]
    \centering
    \includegraphics[width=\linewidth]{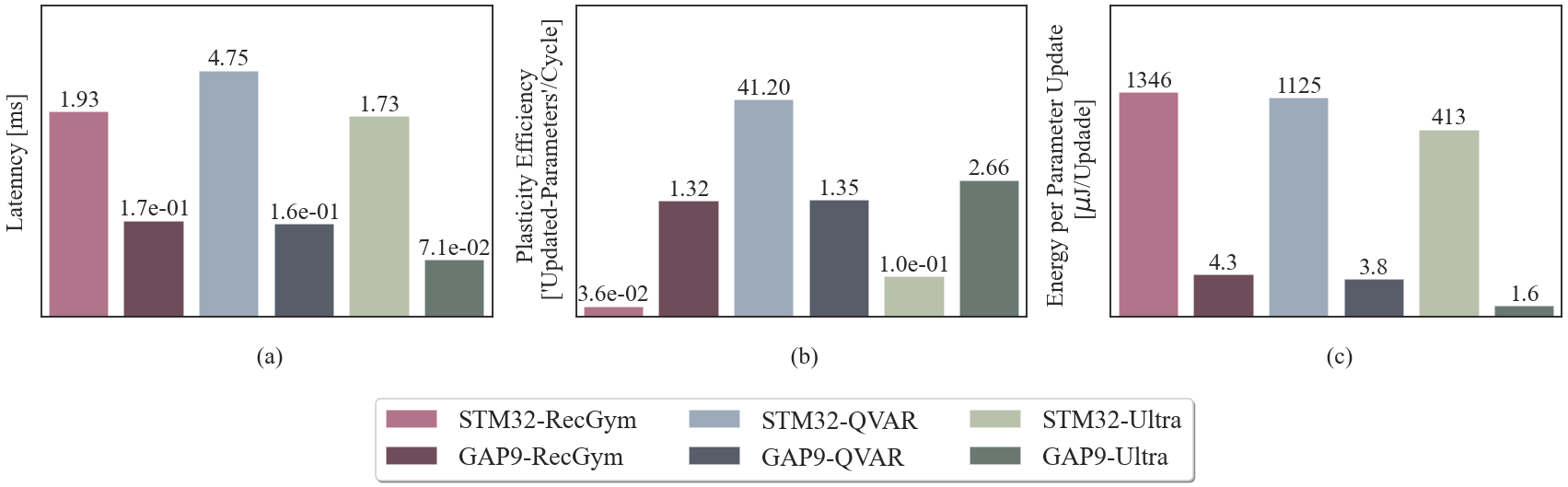}
    \caption{Comparison of the networks' on-device parameter update for three datasets RecGym, Qvar and Ultra. The networks deployed on GAP9 are run with 200MHz while the networks deployed on STM32F756ZG are run with 216MHz.}
    \label{fig:ot_parameter_update}
\end{figure*}

\paragraph{Inference evaluation}
The inference analysis is conducted using the usual metrics latency, inference efficiency ---a metric to assess the target hardware's capability to parallelize MAC operations--- and the energy per inference. Figure \ref{fig:ot_inference}a) presents the number of network parameters, which translates one-to-one into the required memory needed for all the weights. Notably, even though the network for the Ultra dataset is the smallest, they are all within the same order of magnitude and differ in size of $\pm 10$\%. Further, Fig. \ref{fig:ot_inference}(b) and (c) clearly indicate the strength of GAP9 compared to the STM32F7. GAP9 parallelizes the workload 20x fold, and as such is the latency decrease\footnote{The core frequency of the STM32F756ZG is set to 216MHz, while the cluster core frequency of GAP9 is set to 200MHz for this comparison.}. Intriguingly, the energy consumption for running one inference is not 20x smaller. Instead, the energy consumption using the GAP9 instead of STM32F756ZG decreased by more than a factor of 120.

\paragraph{Parameter update evaluation}
The parameter update evaluation is conducted using similar metrics---latency and energy per parameter update---. However, instead of the inference efficiency, we introduce a new metric called \textit{plasticity efficiency}. This metric is used to indicate how many parameters can be updated with one clock cycle. Therefore this metric represents the core's ability to overwrite and as such update the network parameter with respect to one clock cycle. The term plasticity is commonly known in neuroscience and means the brain's ability to learn and adapt, and as such represents the contextual meaning in online training. 
Figure \ref{fig:ot_parameter_update}a) presents the latency needed to update the network's parameters on both target hardware. Interestingly, comparing the parameter update latency with the inference latency of Figure \ref{fig:ot_inference}, parameter update takes approximately one order of magnitude less time than inference. Even though, memory writing is known to be slower, the number of operations is also around one order of magnitude bigger than the amount of network parameters. Further, the plasticity efficiency, shown in Figure \ref{fig:ot_parameter_update}b), as well as the energy per parameter update, shown in Figure \ref{fig:ot_parameter_update}c), indicates GAP9's superior hardware design for on-device AI execution again, even for exotic tasks such as online training. 

%% file: c-conclusion.tex
\section{Conclusion}
In this paper, we propose an on-device transfer learning (ODTL)-based solution for mitigating the user-induced concept drift (UICD) effect that prevails in sensor-based human activity recognition (HAR).
We focus on HAR applications in IoT scenarios, where the computation and memory resources are highly limited.
Specifically, we investigate the UICD effect in three HAR scenarios: human body capacitance (HBC)+IMU-based gym activity recognition (RecGym), electrostatic charge variation+IMU-based hand gesture recognition (QVAR) and 40kHz ultrasonic-based hand gesture recognition (Ultra), as well as the effectiveness of post-deployment ODTL in alleviating the performance loss caused by UICD.
Considering the typical computing hardware for IoT devices, we build our experiments upon two representative MCU-level edge platforms, STM32F756ZG with an ARM Cortex-M7 core, and GAP9 with 8 RISC-V low power cores. For these, we implement optimized on-device training engines as supplements to the native AI deployment tools.
The experimental results reveal varying degrees of UCID loss across the three HAR scenarios. 
However, when user samples are employed for ODTL, a notable improvement in recognition performance is observed. 
Specifically, RecGym, QVAR, and Ultra achieve an increase in recognition performance of 3.73\%, 17.38\%, and 3.70\%, respectively, when compared with the performance achieved through offline training using non-user samples exclusively.
Evaluation of the deployed model in terms of latency and power consumption shows that GAP9 achieves an approximate 20x reduction in both inference and ODTL latency, as well as a remarkable 120x and 280x reduction in inference and ODTL energy consumption, respectively, when compared with STM32F756ZG.
The results prove the advantages of leveraging the latest low-power parallel edge computing device for processing related tasks.
In the future, we plan to conduct further research on the risk management of ODTL to limit the potential performance degradation or crashes associated with ODTL.
In addition, the topic of improving the utilization efficiency of user samples also merits further discussion.

%% file: main.bbl
\begin{thebibliography}{10}
\providecommand{\url}[1]{#1}
\csname url@samestyle\endcsname
\providecommand{\newblock}{\relax}
\providecommand{\bibinfo}[2]{#2}
\providecommand{\BIBentrySTDinterwordspacing}{\spaceskip=0pt\relax}
\providecommand{\BIBentryALTinterwordstretchfactor}{4}
\providecommand{\BIBentryALTinterwordspacing}{\spaceskip=\fontdimen2\font plus
\BIBentryALTinterwordstretchfactor\fontdimen3\font minus \fontdimen4\font\relax}
\providecommand{\BIBforeignlanguage}[2]{{%
\expandafter\ifx\csname l@#1\endcsname\relax
\typeout{** WARNING: IEEEtran.bst: No hyphenation pattern has been}%
\typeout{** loaded for the language `#1'. Using the pattern for}%
\typeout{** the default language instead.}%
\else
\language=\csname l@#1\endcsname
\fi
#2}}
\providecommand{\BIBdecl}{\relax}
\BIBdecl

\bibitem{saleem2023toward}
G.~Saleem, U.~I. Bajwa, and R.~H. Raza, ``Toward human activity recognition: a survey,'' \emph{Neural Computing and Applications}, vol.~35, no.~5, pp. 4145--4182, 2023.

\bibitem{singh2023recent}
R.~Singh, A.~K.~S. Kushwaha, R.~Srivastava \emph{et~al.}, ``Recent trends in human activity recognition--a comparative study,'' \emph{Cognitive Systems Research}, vol.~77, pp. 30--44, 2023.

\bibitem{wang_environment-independent_2021}
\BIBentryALTinterwordspacing
Z.~Wang, S.~Chen, W.~Yang, and Y.~Xu, ``Environment-{Independent} {Wi}-{Fi} {Human} {Activity} {Recognition} with {Adversarial} {Network},'' in \emph{{ICASSP} 2021 - 2021 {IEEE} {International} {Conference} on {Acoustics}, {Speech} and {Signal} {Processing} ({ICASSP})}, Jun. 2021, pp. 3330--3334, iSSN: 2379-190X. [Online]. Available: \url{https://ieeexplore.ieee.org/document/9413590}
\BIBentrySTDinterwordspacing

\bibitem{bharathiraja_real-time_2023}
\BIBentryALTinterwordspacing
N.~Bharathiraja, R.~B. Indhuja, P.~A. Krishnan, S.~Anandhan, and S.~Hariprasad, ``Real-{Time} {Fall} {Detection} using {ESP32} and {AMG8833} {Thermal} {Sensor}: {A} {Non}-{Wearable} {Approach} for {Enhanced} {Safety},'' in \emph{2023 {Second} {International} {Conference} on {Augmented} {Intelligence} and {Sustainable} {Systems} ({ICAISS})}, Aug. 2023, pp. 1732--1736. [Online]. Available: \url{https://ieeexplore.ieee.org/document/10250598}
\BIBentrySTDinterwordspacing

\bibitem{dileep_suspicious_2022}
\BIBentryALTinterwordspacing
A.~S. Dileep, N.~S. S., S.~S., F.~K., and S.~S., ``Suspicious {Human} {Activity} {Recognition} using {2D} {Pose} {Estimation} and {Convolutional} {Neural} {Network},'' in \emph{2022 {International} {Conference} on {Wireless} {Communications} {Signal} {Processing} and {Networking} ({WiSPNET})}, Mar. 2022, pp. 19--23. [Online]. Available: \url{https://ieeexplore.ieee.org/document/9767152}
\BIBentrySTDinterwordspacing

\bibitem{batool_authentication_2020}
\BIBentryALTinterwordspacing
S.~Batool, A.~Hassan, N.~A. Saqib, and M.~A.~K. Khattak, ``Authentication of {Remote} {IoT} {Users} {Based} on {Deeper} {Gait} {Analysis} of {Sensor} {Data},'' \emph{IEEE Access}, vol.~8, pp. 101\,784--101\,796, 2020, conference Name: IEEE Access. [Online]. Available: \url{https://ieeexplore.ieee.org/document/9103069}
\BIBentrySTDinterwordspacing

\bibitem{cassis_intelligent_2022}
\BIBentryALTinterwordspacing
M.~Cassis, ``Intelligent {Sensing}: {Enabling} the {Next} “{Automation} {Age}”,'' in \emph{2022 {IEEE} {International} {Solid}- {State} {Circuits} {Conference} ({ISSCC})}, vol.~65, Feb. 2022, pp. 17--24, iSSN: 2376-8606. [Online]. Available: \url{https://ieeexplore.ieee.org/document/9731666}
\BIBentrySTDinterwordspacing

\bibitem{shi2023robust}
Y.~Shi, L.~Du, X.~Chen, X.~Liao, Z.~Yu, Z.~Li, C.~Wang, and S.~Xue, ``Robust gait recognition based on deep cnns with camera and radar sensor fusion,'' \emph{IEEE Internet of Things Journal}, 2023.

\bibitem{hao2023valerian}
Y.~Hao, B.~Wang, and R.~Zheng, ``Valerian: Invariant feature learning for imu sensor-based human activity recognition in the wild,'' in \emph{Proceedings of the 8th ACM/IEEE Conference on Internet of Things Design and Implementation}, 2023, pp. 66--78.

\bibitem{tamulis2022affective}
{\v{Z}}.~Tamulis, M.~Vasiljevas, R.~Dama{\v{s}}evi{\v{c}}ius, R.~Maskeliunas, and S.~Misra, ``Affective computing for ehealth using low-cost remote internet of things-based emg platform,'' in \emph{Intelligent Internet of Things for Healthcare and Industry}.\hskip 1em plus 0.5em minus 0.4em\relax Springer, 2022, pp. 67--81.

\bibitem{ding2023sparsity}
C.~Ding, L.~Zhang, H.~Chen, H.~Hong, X.~Zhu, and F.~Fioranelli, ``Sparsity-based human activity recognition with pointnet using a portable fmcw radar,'' \emph{IEEE Internet of Things Journal}, 2023.

\bibitem{wei2021real}
W.~Wei, K.~Kurita, J.~Kuang, and A.~Gao, ``Real-time limb motion tracking with a single imu sensor for physical therapy exercises,'' in \emph{2021 43rd Annual International Conference of the IEEE Engineering in Medicine \& Biology Society (EMBC)}.\hskip 1em plus 0.5em minus 0.4em\relax IEEE, 2021, pp. 7152--7157.

\bibitem{ishioka2020single}
H.~Ishioka, X.~Weng, Y.~Man, and K.~Kitani, ``Single camera worker detection, tracking and action recognition in construction site,'' in \emph{ISARC. Proceedings of the International Symposium on Automation and Robotics in Construction}, vol.~37.\hskip 1em plus 0.5em minus 0.4em\relax IAARC Publications, 2020, pp. 653--660.

\bibitem{zhou20244d}
J.~Zhou and J.~Le~Kernec, ``4d radar simulator for human activity recognition,'' \emph{IET Radar, Sonar \& Navigation}, vol.~18, no.~2, pp. 239--255, 2024.

\bibitem{yu2022noninvasive}
C.~Yu, Z.~Xu, K.~Yan, Y.-R. Chien, S.-H. Fang, and H.-C. Wu, ``Noninvasive human activity recognition using millimeter-wave radar,'' \emph{IEEE Systems Journal}, vol.~16, no.~2, pp. 3036--3047, 2022.

\bibitem{abedi2023ai}
H.~Abedi, A.~Ansariyan, P.~P. Morita, A.~Wong, J.~Boger, and G.~Shaker, ``Ai-powered non-contact in-home gait monitoring and activity recognition system based on mm-wave fmcw radar and cloud computing,'' \emph{IEEE Internet of Things Journal}, 2023.

\bibitem{ling_ultragesture_2022}
K.~Ling, H.~Dai, Y.~Liu, A.~X. Liu, W.~Wang, and Q.~Gu, ``{UltraGesture}: {Fine}-{Grained} {Gesture} {Sensing} and {Recognition},'' \emph{IEEE Transactions on Mobile Computing}, vol.~21, no.~7, pp. 2620--2636, Jul. 2022, conference Name: IEEE Transactions on Mobile Computing.

\bibitem{reinschmidt2022realtime}
E.~Reinschmidt, C.~Vogt, and M.~Magno, ``Realtime hand-gesture recognition based on novel charge variation sensor and imu,'' in \emph{2022 IEEE Sensors}.\hskip 1em plus 0.5em minus 0.4em\relax IEEE, 2022, pp. 1--4.

\bibitem{bian2024body}
S.~Bian, M.~Liu, B.~Zhou, P.~Lukowicz, and M.~Magno, ``Body-area capacitive or electric field sensing for human activity recognition and human-computer interaction: A comprehensive survey,'' \emph{Proceedings of the ACM on Interactive, Mobile, Wearable and Ubiquitous Technologies}, vol.~8, no.~1, pp. 1--49, 2024.

\bibitem{wang2023negative}
J.~Wang, T.~Zhu, L.~L. Chen, H.~Ning, and Y.~Wan, ``Negative selection by clustering for contrastive learning in human activity recognition,'' \emph{IEEE Internet of Things Journal}, vol.~10, no.~12, pp. 10\,833--10\,844, 2023.

\bibitem{yang2022efficientfi}
J.~Yang, X.~Chen, H.~Zou, D.~Wang, Q.~Xu, and L.~Xie, ``Efficientfi: Toward large-scale lightweight wifi sensing via csi compression,'' \emph{IEEE Internet of Things Journal}, vol.~9, no.~15, pp. 13\,086--13\,095, 2022.

\bibitem{mcenroe2022survey}
P.~McEnroe, S.~Wang, and M.~Liyanage, ``A survey on the convergence of edge computing and ai for uavs: Opportunities and challenges,'' \emph{IEEE Internet of Things Journal}, vol.~9, no.~17, pp. 15\,435--15\,459, 2022.

\bibitem{wahab2022intrusion}
O.~A. Wahab, ``Intrusion detection in the iot under data and concept drifts: Online deep learning approach,'' \emph{IEEE Internet of Things Journal}, vol.~9, no.~20, pp. 19\,706--19\,716, 2022.

\bibitem{kong2022edge}
X.~Kong, Y.~Wu, H.~Wang, and F.~Xia, ``Edge computing for internet of everything: A survey,'' \emph{IEEE Internet of Things Journal}, vol.~9, no.~23, pp. 23\,472--23\,485, 2022.

\bibitem{ravaglia_tinyml_2021}
\BIBentryALTinterwordspacing
L.~Ravaglia, M.~Rusci, D.~Nadalini, A.~Capotondi, F.~Conti, and L.~Benini, ``A {TinyML} {Platform} for {On}-{Device} {Continual} {Learning} with {Quantized} {Latent} {Replays},'' \emph{IEEE Journal on Emerging and Selected Topics in Circuits and Systems}, vol.~11, no.~4, pp. 789--802, Dec. 2021, arXiv:2110.10486 [cs]. [Online]. Available: \url{http://arxiv.org/abs/2110.10486}
\BIBentrySTDinterwordspacing

\bibitem{ren_tinyol_2021}
H.~Ren, D.~Anicic, and T.~A. Runkler, ``{TinyOL}: {TinyML} with {Online}-{Learning} on {Microcontrollers},'' in \emph{2021 {International} {Joint} {Conference} on {Neural} {Networks} ({IJCNN})}, Jul. 2021, pp. 1--8, iSSN: 2161-4407.

\bibitem{prakash2022iot}
P.~Prakash, J.~Ding, R.~Chen, X.~Qin, M.~Shu, Q.~Cui, Y.~Guo, and M.~Pan, ``Iot device friendly and communication-efficient federated learning via joint model pruning and quantization,'' \emph{IEEE Internet of Things Journal}, vol.~9, no.~15, pp. 13\,638--13\,650, 2022.

\bibitem{lu2023snpf}
Y.~Lu, Z.~Guan, W.~Zhao, M.~Gong, W.~Wang, and K.~Sheng, ``Snpf: Sensitiveness based network pruning framework for efficient edge computing,'' \emph{IEEE Internet of Things Journal}, 2023.

\bibitem{wang2023coopfl}
Z.~Wang, H.~Xu, Y.~Xu, Z.~Jiang, and J.~Liu, ``Coopfl: Accelerating federated learning with dnn partitioning and offloading in heterogeneous edge computing,'' \emph{Computer Networks}, vol. 220, p. 109490, 2023.

\bibitem{choi2023simplification}
K.~Choi, S.~M. Wi, H.~G. Jung, and J.~K. Suhr, ``Simplification of deep neural network-based object detector for real-time edge computing,'' \emph{Sensors}, vol.~23, no.~7, p. 3777, 2023.

\bibitem{MAX78000}
\BIBentryALTinterwordspacing
A.~Devices. Max78000 user guide. Online document. [Online]. Available: \url{https://www.analog.com/media/en/technical-documentation/user-guides/max78000-user-guide.pdf}
\BIBentrySTDinterwordspacing

\bibitem{NDP100}
\BIBentryALTinterwordspacing
Syntiant. Ndp100 neural decision processor. Online document. [Online]. Available: \url{https://static1.squarespace.com/static/6488b0b8150a045d2d112999/t/650b28cfb87ca75617f80776/1695230160161/Syntiant_Product_Brief_NDP100.pdf}
\BIBentrySTDinterwordspacing

\bibitem{Hailo8}
\BIBentryALTinterwordspacing
A.~Devices. Hailo tappas user guide. Online document. [Online]. Available: \url{https://f.hubspotusercontent30.net/hubfs/3383687/TAPPAS%20User%20Guide.pdf}
\BIBentrySTDinterwordspacing

\bibitem{STM32N6}
\BIBentryALTinterwordspacing
STMicroeectronics. Stm32n6: Get a sneak peek at the future of ai-powered mcus. Online document. [Online]. Available: \url{https://blog.st.com/stm32n6/}
\BIBentrySTDinterwordspacing

\bibitem{giordano2022survey}
M.~Giordano, L.~Piccinelli, and M.~Magno, ``Survey and comparison of milliwatts micro controllers for tiny machine learning at the edge,'' in \emph{2022 IEEE 4th International Conference on Artificial Intelligence Circuits and Systems (AICAS)}.\hskip 1em plus 0.5em minus 0.4em\relax IEEE, 2022, pp. 94--97.

\bibitem{moosmann2023ultra}
J.~Moosmann, P.~Bonazzi, Y.~Li, S.~Bian, P.~Mayer, L.~Benini, and M.~Magno, ``Ultra-efficient on-device object detection on ai-integrated smart glasses with tinyissimoyolo,'' \emph{arXiv preprint arXiv:2311.01057}, 2023.

\bibitem{bian2022exploring}
S.~Bian, X.~Wang, T.~Polonelli, and M.~Magno, ``Exploring automatic gym workouts recognition locally on wearable resource-constrained devices,'' in \emph{2022 IEEE 13th International Green and Sustainable Computing Conference (IGSC)}.\hskip 1em plus 0.5em minus 0.4em\relax IEEE, 2022, pp. 1--6.

\bibitem{lattanzi2022exploring}
E.~Lattanzi, M.~Donati, and V.~Freschi, ``Exploring artificial neural networks efficiency in tiny wearable devices for human activity recognition,'' \emph{Sensors}, vol.~22, no.~7, p. 2637, 2022.

\bibitem{incel2023device}
O.~D. Incel and S.~{\"O}. Bursa, ``On-device deep learning for mobile and wearable sensing applications: A review,'' \emph{IEEE Sensors Journal}, vol.~23, no.~6, pp. 5501--5512, 2023.

\bibitem{saha2022machine}
S.~S. Saha, S.~S. Sandha, and M.~Srivastava, ``Machine learning for microcontroller-class hardware: A review,'' \emph{IEEE Sensors Journal}, vol.~22, no.~22, pp. 21\,362--21\,390, 2022.

\bibitem{craighero2023device}
M.~Craighero, D.~Quarantiello, B.~Rossi, D.~Carrera, P.~Fragneto, and G.~Boracchi, ``On-device personalization for human activity recognition on stm32,'' \emph{IEEE Embedded Systems Letters}, 2023.

\bibitem{hayajneh2024tinyml}
A.~M. Hayajneh, M.~Hafeez, S.~A. Zaidi, and D.~McLernon, ``Tinyml empowered transfer learning on the edge,'' \emph{IEEE Open Journal of the Communications Society}, 2024.

\bibitem{ferrari2020personalization}
A.~Ferrari, D.~Micucci, M.~Mobilio, and P.~Napoletano, ``On the personalization of classification models for human activity recognition,'' \emph{IEEE Access}, vol.~8, pp. 32\,066--32\,079, 2020.

\bibitem{an2023transfer}
S.~An, G.~Bhat, S.~Gumussoy, and U.~Ogras, ``Transfer learning for human activity recognition using representational analysis of neural networks,'' \emph{ACM Transactions on Computing for Healthcare}, vol.~4, no.~1, pp. 1--21, 2023.

\bibitem{mathur2020scaling}
A.~Mathur, ``Scaling machine learning systems using domain adaptation,'' Ph.D. dissertation, UCL (University College London), 2020.

\bibitem{kwon2023lifelearner}
Y.~D. Kwon, J.~Chauhan, H.~Jia, S.~I. Venieris, and C.~Mascolo, ``Lifelearner: Hardware-aware meta continual learning system for embedded computing platforms,'' \emph{arXiv preprint arXiv:2311.11420}, 2023.

\bibitem{chowdhary2023sensor}
M.~Chowdhary and S.~S. Saha, ``On-sensor online learning and classification under 8 kb memory,'' in \emph{2023 26th International Conference on Information Fusion (FUSION)}.\hskip 1em plus 0.5em minus 0.4em\relax IEEE, 2023, pp. 1--8.

\bibitem{wang2020accurate}
X.~Wang, M.~Hersche, B.~T{\"o}mekce, B.~Kaya, M.~Magno, and L.~Benini, ``An accurate eegnet-based motor-imagery brain--computer interface for low-power edge computing,'' in \emph{2020 IEEE international symposium on medical measurements and applications (MeMeA)}.\hskip 1em plus 0.5em minus 0.4em\relax IEEE, 2020, pp. 1--6.

\bibitem{craighero_-device_2023}
\BIBentryALTinterwordspacing
M.~Craighero, D.~Quarantiello, B.~Rossi, D.~Carrera, P.~Fragneto, and G.~Boracchi, ``\BIBforeignlanguage{en}{On-{Device} {Personalization} for {Human} {Activity} {Recognition} on {STM32}},'' \emph{\BIBforeignlanguage{en}{IEEE Embedded Systems Letters}}, pp. 1--1, 2023. [Online]. Available: \url{https://ieeexplore.ieee.org/document/10178063/}
\BIBentrySTDinterwordspacing

\bibitem{llisterri_gimenez_-device_2022}
\BIBentryALTinterwordspacing
N.~Llisterri~Giménez, M.~Monfort~Grau, R.~Pueyo~Centelles, and F.~Freitag, ``\BIBforeignlanguage{en}{On-{Device} {Training} of {Machine} {Learning} {Models} on {Microcontrollers} with {Federated} {Learning}},'' \emph{\BIBforeignlanguage{en}{Electronics}}, vol.~11, no.~4, p. 573, Jan. 2022, number: 4 Publisher: Multidisciplinary Digital Publishing Institute. [Online]. Available: \url{https://www.mdpi.com/2079-9292/11/4/573}
\BIBentrySTDinterwordspacing

\bibitem{nadalini2023reduced}
D.~Nadalini, M.~Rusci, L.~Benini, and F.~Conti, ``Reduced precision floating-point optimization for deep neural network on-device learning on microcontrollers,'' \emph{Future Generation Computer Systems}, vol. 149, pp. 212--226, 2023.

\bibitem{fu2020sensing}
B.~Fu, N.~Damer, F.~Kirchbuchner, and A.~Kuijper, ``Sensing technology for human activity recognition: A comprehensive survey,'' \emph{Ieee Access}, vol.~8, pp. 83\,791--83\,820, 2020.

\bibitem{bian2022state}
S.~Bian, M.~Liu, B.~Zhou, and P.~Lukowicz, ``The state-of-the-art sensing techniques in human activity recognition: A survey,'' \emph{Sensors}, vol.~22, no.~12, p. 4596, 2022.

\bibitem{gowda_human_2017}
\BIBentryALTinterwordspacing
S.~N. Gowda, ``Human {Activity} {Recognition} {Using} {Combinatorial} {Deep} {Belief} {Networks},'' in \emph{2017 {IEEE} {Conference} on {Computer} {Vision} and {Pattern} {Recognition} {Workshops} ({CVPRW})}, Jul. 2017, pp. 1589--1594, iSSN: 2160-7516. [Online]. Available: \url{https://ieeexplore.ieee.org/document/8014937}
\BIBentrySTDinterwordspacing

\bibitem{agarwal_weighted_2015}
\BIBentryALTinterwordspacing
I.~Agarwal, A.~K.~S. Kushwaha, and R.~Srivastava, ``Weighted {Fast} {Dynamic} {Time} {Warping} based multi-view human activity recognition using a {RGB}-{D} sensor,'' in \emph{2015 {Fifth} {National} {Conference} on {Computer} {Vision}, {Pattern} {Recognition}, {Image} {Processing} and {Graphics} ({NCVPRIPG})}, Dec. 2015, pp. 1--4. [Online]. Available: \url{https://ieeexplore.ieee.org/document/7490046}
\BIBentrySTDinterwordspacing

\bibitem{beddiar_vision-based_2020}
\BIBentryALTinterwordspacing
D.~R. Beddiar, B.~Nini, M.~Sabokrou, and A.~Hadid, ``\BIBforeignlanguage{en}{Vision-based human activity recognition: a survey},'' \emph{\BIBforeignlanguage{en}{Multimedia Tools and Applications}}, vol.~79, no.~41, pp. 30\,509--30\,555, Nov. 2020. [Online]. Available: \url{https://doi.org/10.1007/s11042-020-09004-3}
\BIBentrySTDinterwordspacing

\bibitem{suh_worker_2023}
\BIBentryALTinterwordspacing
S.~Suh, V.~F. Rey, S.~Bian, Y.-C. Huang, J.~M. Rožanec, H.~T. Ghinani, B.~Zhou, and P.~Lukowicz, ``Worker {Activity} {Recognition} in {Manufacturing} {Line} {Using} {Near}-body {Electric} {Field},'' \emph{IEEE Internet of Things Journal}, pp. 1--1, 2023, conference Name: IEEE Internet of Things Journal. [Online]. Available: \url{https://ieeexplore.ieee.org/abstract/document/10308956}
\BIBentrySTDinterwordspacing

\bibitem{pearce2023multi}
A.~Pearce, J.~A. Zhang, R.~Xu, and K.~Wu, ``Multi-object tracking with mmwave radar: A review,'' \emph{Electronics}, vol.~12, no.~2, p. 308, 2023.

\bibitem{chioccarello2023forte}
S.~Chioccarello, A.~Slu{\"y}ters, A.~Testolin, J.~Vanderdonckt, and S.~Lambot, ``Forte: Few samples for recognizing hand gestures with a smartphone-attached radar,'' \emph{Proceedings of the ACM on Human-Computer Interaction}, vol.~7, no. EICS, pp. 1--25, 2023.

\bibitem{liu2024unifi}
Y.~Liu, A.~Yu, L.~Wang, B.~Guo, Y.~Li, E.~Yi, and D.~Zhang, ``Unifi: A unified framework for generalizable gesture recognition with wi-fi signals using consistency-guided multi-view networks,'' \emph{Proceedings of the ACM on Interactive, Mobile, Wearable and Ubiquitous Technologies}, vol.~7, no.~4, pp. 1--29, 2024.

\bibitem{lian2023echosensor}
J.~Lian, C.~Du, J.~Lou, L.~Chen, and X.~Yuan, ``Echosensor: Fine-grained ultrasonic sensing for smart home intrusion detection,'' \emph{ACM Transactions on Sensor Networks}, vol.~20, no.~1, pp. 1--24, 2023.

\bibitem{hussain2023low}
A.~Hussain, S.~U. Khan, N.~Khan, I.~Rida, M.~Alharbi, and S.~W. Baik, ``Low-light aware framework for human activity recognition via optimized dual stream parallel network,'' \emph{Alexandria Engineering Journal}, vol.~74, pp. 569--583, 2023.

\bibitem{zhou_efficient_2020}
F.~Zhou, X.~Li, and Z.~Wang, ``Efficient {High} {Cross}-{User} {Recognition} {Rate} {Ultrasonic} {Hand} {Gesture} {Recognition} {System},'' \emph{IEEE Sensors Journal}, vol.~20, no.~22, pp. 13\,501--13\,510, Nov. 2020, conference Name: IEEE Sensors Journal.

\bibitem{yang2023ultradigit}
M.~Yang, J.~Zhang, and X.~Wang, ``Ultradigit: An ultrasound signal-based in-air digit input system via transfer learning,'' \emph{IET Radar, Sonar \& Navigation}, vol.~17, no.~11, pp. 1674--1687, 2023.

\bibitem{wang2022faceori}
Y.~Wang, J.~Ding, I.~Chatterjee, F.~Salemi~Parizi, Y.~Zhuang, Y.~Yan, S.~Patel, and Y.~Shi, ``Faceori: Tracking head position and orientation using ultrasonic ranging on earphones,'' in \emph{Proceedings of the 2022 CHI Conference on Human Factors in Computing Systems}, 2022, pp. 1--12.

\bibitem{bian2022contribution}
S.~Bian, V.~F. Rey, S.~Yuan, and P.~Lukowicz, ``The contribution of human body capacitance/body-area electric field to individual and collaborative activity recognition,'' \emph{arXiv preprint arXiv:2210.14794}, 2022.

\bibitem{bian2022using}
S.~Bian, S.~Yuan, V.~F. Rey, and P.~Lukowicz, ``Using human body capacitance sensing to monitor leg motion dominated activities with a wrist worn device,'' in \emph{Sensor-and Video-Based Activity and Behavior Computing: Proceedings of 3rd International Conference on Activity and Behavior Computing (ABC 2021)}.\hskip 1em plus 0.5em minus 0.4em\relax Springer, 2022, pp. 81--94.

\bibitem{bian2019wrist}
S.~Bian, V.~F. Rey, J.~Younas, and P.~Lukowicz, ``Wrist-worn capacitive sensor for activity and physical collaboration recognition,'' in \emph{2019 IEEE International Conference on Pervasive Computing and Communications Workshops (PerCom Workshops)}.\hskip 1em plus 0.5em minus 0.4em\relax IEEE, 2019, pp. 261--266.

\bibitem{dheman2022cardiac}
K.~Dheman, D.~Werder, and M.~Magno, ``Cardiac monitoring with novel low power sensors measuring upper thoracic electrostatic charge variation for long lasting wearable devices,'' in \emph{2022 18th International Conference on Wireless and Mobile Computing, Networking and Communications (WiMob)}.\hskip 1em plus 0.5em minus 0.4em\relax IEEE, 2022, pp. 154--159.

\bibitem{manoni2022long}
A.~Manoni, A.~Gumiero, A.~Zampogna, C.~Ciarlo, L.~Panetta, A.~Suppa, L.~Della~Torre, and F.~Irrera, ``Long-term polygraphic monitoring through mems and charge transfer for low-power wearable applications,'' \emph{Sensors}, vol.~22, no.~7, p. 2566, 2022.

\bibitem{de_vita_-ff_2023}
\BIBentryALTinterwordspacing
F.~De~Vita, R.~M.~A. Nawaiseh, D.~Bruneo, V.~Tomaselli, M.~Lattuada, and M.~Falchetto, ``µ-{FF}: {On}-{Device} {Forward}-{Forward} {Training} {Algorithm} for {Microcontrollers},'' in \emph{2023 {IEEE} {International} {Conference} on {Smart} {Computing} ({SMARTCOMP})}, Jun. 2023, pp. 49--56, iSSN: 2693-8340. [Online]. Available: \url{https://ieeexplore.ieee.org/document/10207585}
\BIBentrySTDinterwordspacing

\bibitem{pau_automated_2022}
\BIBentryALTinterwordspacing
D.~Pau and P.~K. Ambrose, ``\BIBforeignlanguage{en}{Automated {Neural} and {On}-{Device} {Learning} for {Micro} {Controllers}},'' in \emph{\BIBforeignlanguage{en}{2022 {IEEE} 21st {Mediterranean} {Electrotechnical} {Conference} ({MELECON})}}.\hskip 1em plus 0.5em minus 0.4em\relax Palermo, Italy: IEEE, Jun. 2022, pp. 758--763. [Online]. Available: \url{https://ieeexplore.ieee.org/document/9843050/}
\BIBentrySTDinterwordspacing

\bibitem{disabato_incremental_2020}
\BIBentryALTinterwordspacing
S.~Disabato and M.~Roveri, ``Incremental {On}-{Device} {Tiny} {Machine} {Learning},'' in \emph{Proceedings of the 2nd {International} {Workshop} on {Challenges} in {Artificial} {Intelligence} and {Machine} {Learning} for {Internet} of {Things}}, ser. {AIChallengeIoT} '20.\hskip 1em plus 0.5em minus 0.4em\relax New York, NY, USA: Association for Computing Machinery, Nov. 2020, pp. 7--13. [Online]. Available: \url{https://dl.acm.org/doi/10.1145/3417313.3429378}
\BIBentrySTDinterwordspacing

\bibitem{pau_online_2021}
\BIBentryALTinterwordspacing
D.~Pau, A.~Khiari, and D.~Denaro, ``Online learning on tiny micro-controllers for anomaly detection in water distribution systems,'' in \emph{2021 {IEEE} 11th {International} {Conference} on {Consumer} {Electronics} ({ICCE}-{Berlin})}, Nov. 2021, pp. 1--6, iSSN: 2166-6822. [Online]. Available: \url{https://ieeexplore.ieee.org/document/9720009}
\BIBentrySTDinterwordspacing

\bibitem{abdennadher_fixed_2021}
\BIBentryALTinterwordspacing
N.~Abdennadher, D.~Pau, and A.~Bruna, ``Fixed complexity tiny reservoir heterogeneous network for on-line {ECG} learning of anomalies,'' in \emph{2021 {IEEE} 10th {Global} {Conference} on {Consumer} {Electronics} ({GCCE})}, Oct. 2021, pp. 233--237, iSSN: 2378-8143. [Online]. Available: \url{https://ieeexplore.ieee.org/document/9622022}
\BIBentrySTDinterwordspacing

\bibitem{lin_-device_2022}
\BIBentryALTinterwordspacing
J.~Lin, L.~Zhu, W.-M. Chen, W.-C. Wang, C.~Gan, and S.~Han, ``\BIBforeignlanguage{en}{On-{Device} {Training} {Under} {256KB} {Memory}},'' \emph{\BIBforeignlanguage{en}{Advances in Neural Information Processing Systems}}, vol.~35, pp. 22\,941--22\,954, Dec. 2022. [Online]. Available: \url{https://proceedings.neurips.cc/paper_files/paper/2022/hash/90c56c77c6df45fc8e556a096b7a2b2e-Abstract-Conference.html}
\BIBentrySTDinterwordspacing

\bibitem{khan_dacapo_2023}
\BIBentryALTinterwordspacing
O.~Khan, G.~Park, and E.~Seo, ``{DaCapo}: {An} {On}-{Device} {Learning} {Scheme} for {Memory}-{Constrained} {Embedded} {Systems},'' \emph{ACM Transactions on Embedded Computing Systems}, vol.~22, no.~5s, pp. 142:1--142:23, 2023. [Online]. Available: \url{https://dl.acm.org/doi/10.1145/3609121}
\BIBentrySTDinterwordspacing

\bibitem{kwon_tinytrain_2023}
\BIBentryALTinterwordspacing
Y.~D. Kwon, R.~Li, S.~I. Venieris, J.~Chauhan, N.~D. Lane, and C.~Mascolo, ``{TinyTrain}: {Deep} {Neural} {Network} {Training} at the {Extreme} {Edge},'' Jul. 2023, arXiv:2307.09988 [cs]. [Online]. Available: \url{http://arxiv.org/abs/2307.09988}
\BIBentrySTDinterwordspacing

\bibitem{rossi_44_2021}
\BIBentryALTinterwordspacing
D.~Rossi, F.~Conti, M.~Eggiman, S.~Mach, A.~D. Mauro, M.~Guermandi, G.~Tagliavini, A.~Pullini, I.~Loi, J.~Chen, E.~Flamand, and L.~Benini, ``4.4 {A} 1.{3TOPS}/{W} @ {32GOPS} {Fully} {Integrated} 10-{Core} {SoC} for {IoT} {End}-{Nodes} with 1.{7$\mu$W} {Cognitive} {Wake}-{Up} {From} {MRAM}-{Based} {State}-{Retentive} {Sleep} {Mode},'' in \emph{2021 {IEEE} {International} {Solid}- {State} {Circuits} {Conference} ({ISSCC})}, vol.~64, Feb. 2021, pp. 60--62, iSSN: 2376-8606. [Online]. Available: \url{https://ieeexplore.ieee.org/document/9365939}
\BIBentrySTDinterwordspacing

\bibitem{kopparapu_tinyfedtl_2022}
\BIBentryALTinterwordspacing
K.~Kopparapu, E.~Lin, J.~G. Breslin, and B.~Sudharsan, ``{TinyFedTL}: {Federated} {Transfer} {Learning} on {Ubiquitous} {Tiny} {IoT} {Devices},'' in \emph{2022 {IEEE} {International} {Conference} on {Pervasive} {Computing} and {Communications} {Workshops} and other {Affiliated} {Events} ({PerCom} {Workshops})}, Mar. 2022, pp. 79--81. [Online]. Available: \url{https://ieeexplore.ieee.org/abstract/document/9767250}
\BIBentrySTDinterwordspacing

\bibitem{avi_incremental_2022}
\BIBentryALTinterwordspacing
A.~Avi, A.~Albanese, and D.~Brunelli, ``\BIBforeignlanguage{en}{Incremental {Online} {Learning} {Algorithms} {Comparison} for {Gesture} and {Visual} {Smart} {Sensors}},'' in \emph{\BIBforeignlanguage{en}{2022 {International} {Joint} {Conference} on {Neural} {Networks} ({IJCNN})}}.\hskip 1em plus 0.5em minus 0.4em\relax Padua, Italy: IEEE, Jul. 2022, pp. 1--8. [Online]. Available: \url{https://ieeexplore.ieee.org/document/9892356/}
\BIBentrySTDinterwordspacing

\bibitem{RecGym}
\BIBentryALTinterwordspacing
kaggle.com, ``Recgym: Gym workouts data set.'' [Online]. Available: \url{https://www.kaggle.com/datasets/zhaxidelebsz/10-gym-exercises-with-615-abstracted-features/settings}
\BIBentrySTDinterwordspacing

\bibitem{kangpx_40khz-ultrasonicdhgr-onlinesemi_2023}
\BIBentryALTinterwordspacing
kangpx, ``40khz-{ultrasonicDHGR}-{onlineSemi},'' Jan. 2023, original-date: 2022-12-14T08:54:14Z. [Online]. Available: \url{https://github.com/kangpx/40khz-ultrasonicDHGR-onlineSemi}
\BIBentrySTDinterwordspacing

\bibitem{bian2019passive}
S.~Bian, V.~F. Rey, P.~Hevesi, and P.~Lukowicz, ``Passive capacitive based approach for full body gym workout recognition and counting,'' in \emph{2019 IEEE International Conference on Pervasive Computing and Communications (PerCom}.\hskip 1em plus 0.5em minus 0.4em\relax IEEE, 2019, pp. 1--10.

\end{thebibliography}
